\documentclass[longauth]{aa} 

\usepackage[utf8]{inputenc}
\usepackage[varg]{txfonts}
\usepackage[breaklinks, colorlinks, citecolor=blue, linkcolor=blue]{hyperref}
\usepackage[usenames, dvipsnames]{color}
\usepackage{amsmath}
\usepackage{graphicx}
\usepackage{caption}
\usepackage{subcaption}
\usepackage[english]{babel}
\usepackage{siunitx}
\usepackage{float}
\usepackage{url}
\usepackage{longtable}
\usepackage{fancyhdr}
\usepackage{natbib}
\usepackage[normalem]{ulem}
\usepackage[version=4]{mhchem}

\makeatletter

\def\instrefs#1{{\def\scsep{\def\scsep{,}}\@for\w:=#1\do{\scsep\ref{inst:\w}}}}

\renewcommand{\inst}[1]{\unskip$^{\instrefs{#1}}$}

\let\orgautoref\autoref
\renewcommand{\autoref}
        {\def\equationautorefname{Eq.}
         \def\figureautorefname{Fig.}
         \def\sectionautorefname{Sect.}
         \def\subsectionautorefname{Sect.}
         \def\subsubsectionautorefname{Sect.}
         \orgautoref}

\renewcommand*\aa@pageof{, page \thepage{} of \pageref*{LastPage}}

\newcommand{\tess}{TESS~}

\makeatother

\title{Mass and density of the transiting hot and rocky super-Earth LHS~1478~b (TOI-1640~b)}

\author{
M.\,G.~Soto\inst{qmul}\thanks{E-mail: \tt{m.soto@qmul.ac.uk}}
\and G.~Anglada-Escud\'e\inst{qmul,ice,ieec}
\and S.~Dreizler\inst{iag}
\and K.~Molaverdikhani\inst{lsw,mpia}
\and J.~Kemmer\inst{lsw}
\and C.~Rodr\'iguez-L\'opez\inst{iaa}
\and J.~Lillo-Box\inst{cabesac}
\and E.~Pall\'e\inst{iac,ull}
\and N.~Espinoza\inst{stsci}
\and J.\,A.~Caballero\inst{cabesac}
\and A.~Quirrenbach\inst{lsw}
\and I.~Ribas\inst{ice,ieec}
\and A.~Reiners\inst{iag}
\and N.~Narita\inst{ikis,ijsta,iabc,iac}
\and T.~Hirano\inst{iuteps,iabc,inao}
\and P.\,J.~Amado\inst{iaa}
\and V.\,J.\,S.~B\'ejar\inst{iac,ull}
\and P.~Bluhm\inst{lsw}
\and C.~J.~Burke\inst{mitkavli}
\and D.\,A.~Caldwell\inst{seti,ames}
\and D.~Charbonneau\inst{cfa} 
\and R.~Cloutier\inst{cfa} 
\and K.\,A.~Collins\inst{cfa} 
\and M.~Cort\'es-Contreras\inst{cabesac}
\and E.~Girardin\inst{gpo} 
\and P.~Guerra\inst{oaa} 
\and H.~Harakawa\inst{isubaru}
\and A.\,P.~Hatzes\inst{tls}
\and J.~Irwin\inst{cfa} 
\and J.\,M.~Jenkins\inst{ames}
\and E.~Jensen\inst{swarthmore} 
\and K.~Kawauchi\inst{iuteps}
\and T.~Kotani\inst{iabc,inao,idas}
\and T.~Kudo\inst{isubaru}
\and M.~Kunimoto\inst{mitkavli}
\and M.~Kuzuhara\inst{iabc,inao}
\and D.\,W.~Latham\inst{cfa}
\and D.~Montes\inst{ucm}
\and J.\,C.~Morales\inst{ice,ieec}
\and M.~Mori\inst{iutda}
\and R.~P.~Nelson\inst{qmul}
\and M.~Omiya\inst{iabc,inao}
\and S.~Pedraz\inst{caha}
\and V.\,M.~Passegger\inst{hs,uok}
\and B.\,V.~Rackham\inst{mit}
\and A.~Rudat\inst{mitkavli}
\and J.~E.~Schlieder\inst{goddard}
\and P.~Sch\"ofer\inst{iag}
\and A.~Schweitzer\inst{hs}
\and A.~Selezneva\inst{oaa}
\and C.~Stockdale\inst{hazelwood} 
\and M.~Tamura\inst{iutda,iabc,inao}
\and T.~Trifonov\inst{mpia}
\and R.~Vanderspek\inst{mitkavli}
\and D.~Watanabe\inst{pd}
}

\institute{
\label{inst:qmul}School of Physics and Astronomy, Queen Mary University London, 327 Mile End Road, London E1 4NS, UK
\and
\label{inst:ice}Institut de Ci\`encies de l’Espai (ICE, CSIC), Campus UAB, Can Magrans s/n, 08193 Bellaterra, Spain
\and
\label{inst:ieec}Institut d’Estudis Espacials de Catalunya (IEEC), 08034 Barcelona, Spain
\and
\label{inst:iag}Institut f\"ur Astrophysik, Georg-August-Universit\"at, Friedrich-Hund-Platz 1, 37077 G\"ottingen, Germany
\and
\label{inst:lsw}Landessternwarte, Zentrum f\"ur Astronomie der Universit\"at Heidelberg, K\"onigstuhl 12, 69117 Heidelberg, Germany
\and
\label{inst:mpia}Max-Planck-Institut f\"ur Astronomie, K\"onigstuhl 17, 69117 Heidelberg, Germany
\and
\label{inst:iaa}Instituto de Astrof\'isica de Andaluc\'ia (IAA-CSIC), Glorieta de la Astronom\'ia s/n, 18008 Granada, Spain
\and 
\label{inst:cabesac}Centro de Astrobiolog\'ia (CSIC-INTA), ESAC, Camino bajo del castillo s/n, 28692 Villanueva de la Ca\~nada, Madrid, Spain
\and
\label{inst:iac}Instituto de Astrof\'isica de Canarias, 38205 La Laguna, Tenerife, Spain
\and 
\label{inst:ull}Departamento de Astrof\'isica, Universidad de La Laguna, 38206 La Laguna, Tenerife, Spain
\and
\label{inst:stsci}Space Telescope Science Institute, 3700 San Martin Drive, Baltimore, MD 21218, USA
\and
\label{inst:ikis}Komaba Institute for Science, The University of Tokyo, 3-8-1 Komaba, Meguro, Tokyo 153-8902, Japan
\and
\label{inst:ijsta}Japan Science and Technology Agency, PRESTO, 3-8-1 Komaba, Meguro, Tokyo 153-8902, Japan
\and
\label{inst:iabc}Astrobiology Center, 2-21-1 Osawa, Mitaka, Tokyo 181-8588, Japan 
\and
\label{inst:iuteps}Department of Earth and Planetary Science, The University of Tokyo, Tokyo, Japan
\and
\label{inst:inao}National Astronomical Observatory of Japan, 2-21-1 Osawa, Mitaka, Tokyo 181-8588, Japan 
\and
\label{inst:seti}SETI Institute, Mountain View, CA 94043, USA
\and
\label{inst:ames}NASA Ames Research Center, Moffett Field, CA 94035, USA
\and
\label{inst:cfa}Center for Astrophysics | Harvard \& Smithsonian, 60 Garden St, Cambridge, MA, 02138, USA
\and
\label{inst:gpo}Grand Pra Observatory, Switzerland
\and
\label{inst:oaa}Observatori Astronòmic Albanyà, Camí de Bassegoda S/N, Albanyà 17733, Girona, Spain
\and
\label{inst:isubaru}Subaru Telescope, 650 N. Aohoku Place, Hilo, HI 96720, USA 
\and
\label{inst:tls}Th\"uringer Landessternwarte Tautenburg, Sternwarte 5, 07778 Tautenburg, Germany
\and
\label{inst:swarthmore}Deptartment of Physics \& Astronomy, Swarthmore College, Swarthmore PA 19081, USA
\and
\label{inst:idas}Department of Astronomy, School of Science, The Graduate University for Advanced Studies (SOKENDAI), 2-21-1 Osawa, Mitaka, Tokyo, Japan
\and
\label{inst:mitkavli}MIT Kavli Institute for Astrophysics and Space Research, Massachusetts Institute of Technology, Cambridge, MA 02139, USA
\and
\label{inst:ucm}Departamento de F\'{i}sica de la Tierra y Astrof\'{i}sica and IPARCOS-UCM (Instituto de F\'{i}sica de Part\'{i}culas y del Cosmos de la UCM), Facultad de Ciencias F\'{i}sicas, Universidad Complutense de Madrid, 28040, Madrid, Spain
\and
\label{inst:iutda}Department of Astronomy, The University of Tokyo, 7-3-1 Hongo, Bunkyo-ku, Tokyo 113-0033, Japan   
\and
\label{inst:caha}Centro Astron\'omico Hispano-Alem\'an, Observatorio de Calar Alto, 04550 G\'ergal, Almer\'ia, Spain
\and
\label{inst:hs}Hamburger Sternwarte, Universit\"at Hamburg, Gojenbergsweg 112, 21029 Hamburg, Germany
\and
\label{inst:uok}Homer L. Dodge Department of Physics and Astronomy, University of Oklahoma, 440 West Brooks Street, Norman, OK 73019, USA
\and
\label{inst:mit}Department of Earth, Atmospheric and Planetary Sciences, Massachusetts Institute of Technology, Cambridge, MA 02139, USA
\and
\label{inst:goddard}NASA Goddard Space Flight Center, 8800 Greenbelt Rd, Greenbelt, MD 20771, USA
\and
\label{inst:hazelwood}Hazelwood Observatory, Australia
\and
\label{inst:pd}Planetary Discoveries, Fredericksburg, VA 22405, USA
}

\date{Received 19 February 2021 / Accepted dd Month 2021}

\abstract{One of the main objectives of the {\em Transiting Exoplanet Survey Satellite} ({TESS}) mission is the discovery of small rocky planets around relatively bright nearby stars. 
Here, we report the discovery and characterization of the transiting super-Earth planet orbiting LHS~1478 (TOI-1640). 
The star is an inactive red dwarf ($J \sim 9.6$\,mag and spectral type m3\,V) with mass and radius estimates of $0.20\pm0.01$\,$M_{\odot}$ and $0.25\pm0.01$\,$R_{\odot}$, respectively, and an effective temperature of $3381\pm54$\,K.
It was observed by \tess in four sectors. These data revealed a transit-like feature with a period of 1.949 days. 
We combined the TESS data with three ground-based transit measurements, 57 radial velocity (RV) measurements from CARMENES, and 13 RV measurements from IRD, determining that the signal is produced by a planet with a mass of $2.33^{+0.20}_{-0.20}$\,$M_{\oplus}$ and a radius of $1.24^{+0.05}_{-0.05}$\,$R_{\oplus}$. The resulting bulk density of this planet is 6.67\,g\,cm$^{-3}$, which is consistent with a rocky planet with an Fe- and MgSiO$_3$-dominated composition.  
Although the planet would be too hot to sustain liquid water on its surface (its equilibrium temperature is about $\sim$595\,K, suggesting a Venus-like atmosphere), spectroscopic metrics based on the capabilities of the forthcoming {\em James Webb Space Telescope} and the fact that the host star is rather inactive indicate that this is one of the most favorable known rocky exoplanets for atmospheric characterization.}


\begin{document} 

\maketitle
%

\section{Introduction}

Stars smaller than the Sun offer a number of advantages as to the detection of exoplanets with both the Doppler \citep[e.g., Proxima~b;][]{anglada:2016} and photometric transit methods \citep[e.g., the TRAPPIST-1 system;][]{gillon:2017}. 
Both are indirect methods that rely on the planet imprinting a signal onto the star light. 
The Doppler technique measures the radial velocity (RV) of the star by measuring to high accuracy the wavelength shift of numerous absorption features in the stellar spectrum. 
The amplitude of the signal is larger for a larger planet-star mass ratio and for shorter orbital periods. The transit method relies on measuring the light blocked by the planet as it crosses the stellar disk. 
In this case, the signal is proportional to the planet-star 
area ratio, and it repeats once per orbital period. 
With both methods, shorter-period planets are, therefore, more easily detectable. 

Exoplanet surveys with the Doppler technique are typically conducted with ground-based high-resolution spectrometers ($R = \lambda/\delta \lambda >$ 50\,000) that are kept in a very stable environment and are calibrated against spectral features of a reference source measured in the laboratory. 
This is the case for the CARMENES spectrometer \citep{Quirrenbach:2016}, which was specifically designed to obtain maximal precision on red dwarf stars. 
The CARMENES survey has led to numerous new exoplanet discoveries in the super-Earth to Earth-mass regime, in hot to warm temperate orbits \citep[e.g.,][]{luque2018,zechmeister:2019,stock2020}. 
Several of these planets have been detected both in transit and by  RV \citep{luque:2019,dreizler2020,kemmer2020,nowak2020,bluhm2020}.
On the other hand, the NASA {\em Transiting Exoplanet Survey Satellite} (TESS) mission (\citealt{ricker:2015}) has been surveying most of the sky for signals of transiting planets, with the goal of detecting those that should enable a more straightforward atmospheric characterization. 
To date, \tess has revealed numerous exoplanet candidates that are transiting nearby M dwarfs \citep[to name a few]{astudillo2020, Gan2020, Kanodia2020, Trifonov2021}.

One key element in the characterization and study of transiting planets is their confirmation using Doppler spectroscopy, which in turn produces a measurement of their masses, allowing one to derive their mean bulk densities and put constrains on their compositions. Finally, the atmospheres of transiting planets can be studied by performing spectroscopic measurements during transits (the planet blocks more light at certain wavelengths where the molecules in its atmosphere deter more light, producing deeper transits) and secondary eclipses (the spectrum of thermal emission of the planet is also affected by the presence of absorbing molecules, producing a shallower dip in the light curve when the planet goes behind the star). Among these, and thanks to the more favorable radius ratio between the planet and the star, exoplanets transiting red dwarfs are the best ones for atmospheric characterization. The measured spectrum of an exoplanet can be affected by the presence of spots and active regions in the visible part of the star during transits \citep{Rackman2018}: The less active the star is, the cleaner and easier it is to interpret the measured spectrum of a planet.
Due to its properties (transiting, Doppler signal in the few m\,s$^{-1}$ regime) and its host star (small, relatively nearby, and low stellar activity), LHS~1478 \,b satisfies all the favorable conditions for becoming a prime target for the characterization of rocky terrestrial planets.

In this paper, we validate the transiting exoplanet LHS~1478~b (TOI-1640~b). We provide an overview of the measurements with photometry (initial detection, stellar activity, orbital period, and planet size), imaging (validation against false positives), and RV (confirmation, stellar activity, and measurement of the planet mass) in Sect.~\ref{sec:data}. 
Stellar parameters are presented in Sect.~\ref{sec:stellar_params}, and a joint analysis to constrain the planet properties is given in Sect.~\ref{sec:joint-fit}. We discuss the results in the context of terrestrial planet candidates and further follow-up in Sect.~\ref{sec:discussion} and summarize our work in Sect.~\ref{sec:conclusions}.

\section{Data}\label{sec:data}

\begin{figure*}
    \centering
    \includegraphics[width=\textwidth]{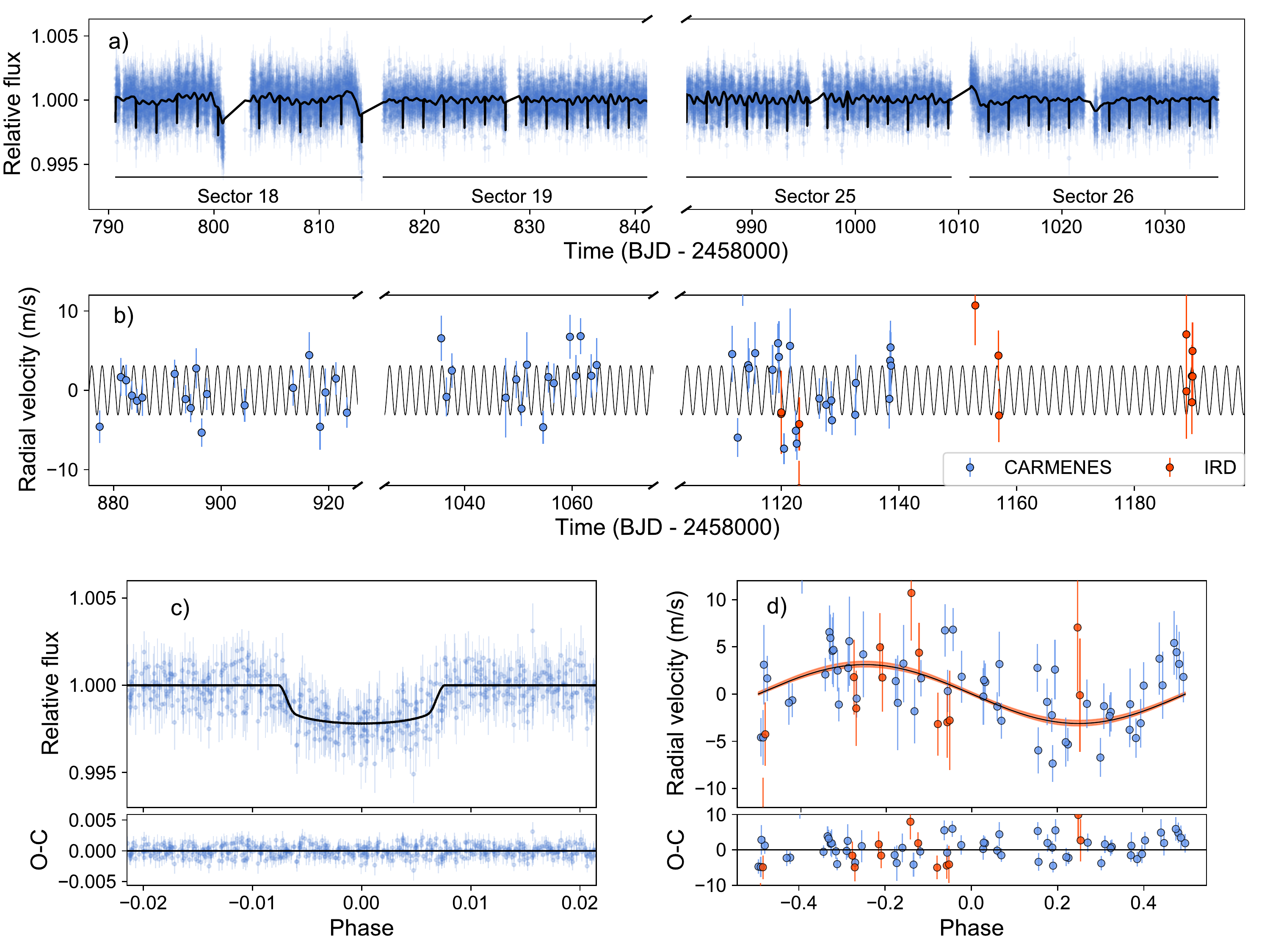}
    \caption{Data and joint fit results of LHS~1478 with \texttt{juliet} (Sect.~\ref{sec:joint-fit}). 
    {\em Top (a)}: \tess photometry, with the black line representing the best transit-plus-GP fit. 
    {\em Middle (b)}: RV data from CARMENES and IRD, with the jitter term added to the uncertainties. The black line is the best fit Keplerian model.
    {\em Bottom left (c)}: Phased-folded \tess data, with the GP component removed. 
    The black line is the best transit fit. 
    The bottom sub-panel shows the residuals from the transit fit.
    {\em Bottom right (d)}: Phased-folded RV data. 
    The black line is the best fit Keplerian model, and the red shaded area represents the 68\,\% CI. The bottom sub-panel shows the residuals after removing the Keplerian fit. 
    }
    \label{fig:joint_fit}
\end{figure*}

\subsection{\tess photometry} \label{sec:tessobs}

LHS~1478 (TOI-1640) was observed by TESS in sectors 18, 19, 25, and 26 of its primary mission. The data were processed through the Science Processing Operations Center \citep[SPOC; ][]{Jenkins2016}, and transiting planet search algorithms \citep{Jenkins2002,Jenkins2010} detected a signal with a period of $1.949521 \pm 0.000007$\,d and a transit depth of $2185 \pm 111$ parts-per-million (ppm) in December 2019 based on the data from sector 18. 
After reviewing the results of  the data validation reports \citep{Twicken2018,Li:DVmodelFit2019}, the TESS Science Office alerted the community of TOI-1640  on 14 January 2020\footnote{\url{https://tess.mit.edu/toi-releases/}}.

We obtained the photometric light curve, corrected for systematics \citep [PDC;][]{Smith2012}, from the Mikulski Archive for Space Telescopes\footnote{\url{https://archive.stsci.edu}} (MAST) using the \texttt{lightkurve}\footnote{\url{https://github.com/KeplerGO/Lightkurve}} package \citep{lightkurve}. 
The data are shown in Fig.~\ref{fig:joint_fit}.
We performed a period search using the Box-Fitting Least Squares \citep[BLS; ][]{BLS} and the Transit Least Squares \citep[TLS; ][]{TLS} algorithms (Fig.~\ref{fig:BLS_TLS}), and we detected with both of them a signal with a period of 1.949\,d and a signal detection efficiency (SDE) $>$ 8 (empirical transit detection threshold from \citealt{aigrain2016}). 
This period is the same as the one reported by the TESS SPOC.

Figure~\ref{fig:TFOV} shows the \tess field of view and aperture size used for each of the four sectors for LHS~1478, generated using \texttt{tpfplotter}\footnote{\url{https://github.com/jlillo/tpfplotter}} \citep{Aller2020}. No sources contaminating the \tess aperture were found in the \textit{Gaia} Early Data Release 3 (EDR3) catalog \citep{GaiaEDR3} down to a magnitude contrast limit of 6\,mag compared to our target. 
Additionally, the {\em Gaia} EDR3 renormalized unit weight error (RUWE) value for this target is 1.26, below the critical value of 1.40, which is an indicator that a source is non-single or has a problematic astrometric solution \citep{Lindegren2020}.

\subsection{High-spatial-resolution imaging}\label{sec:astralux}

\begin{figure}
   \centering
   \includegraphics[width=\hsize]{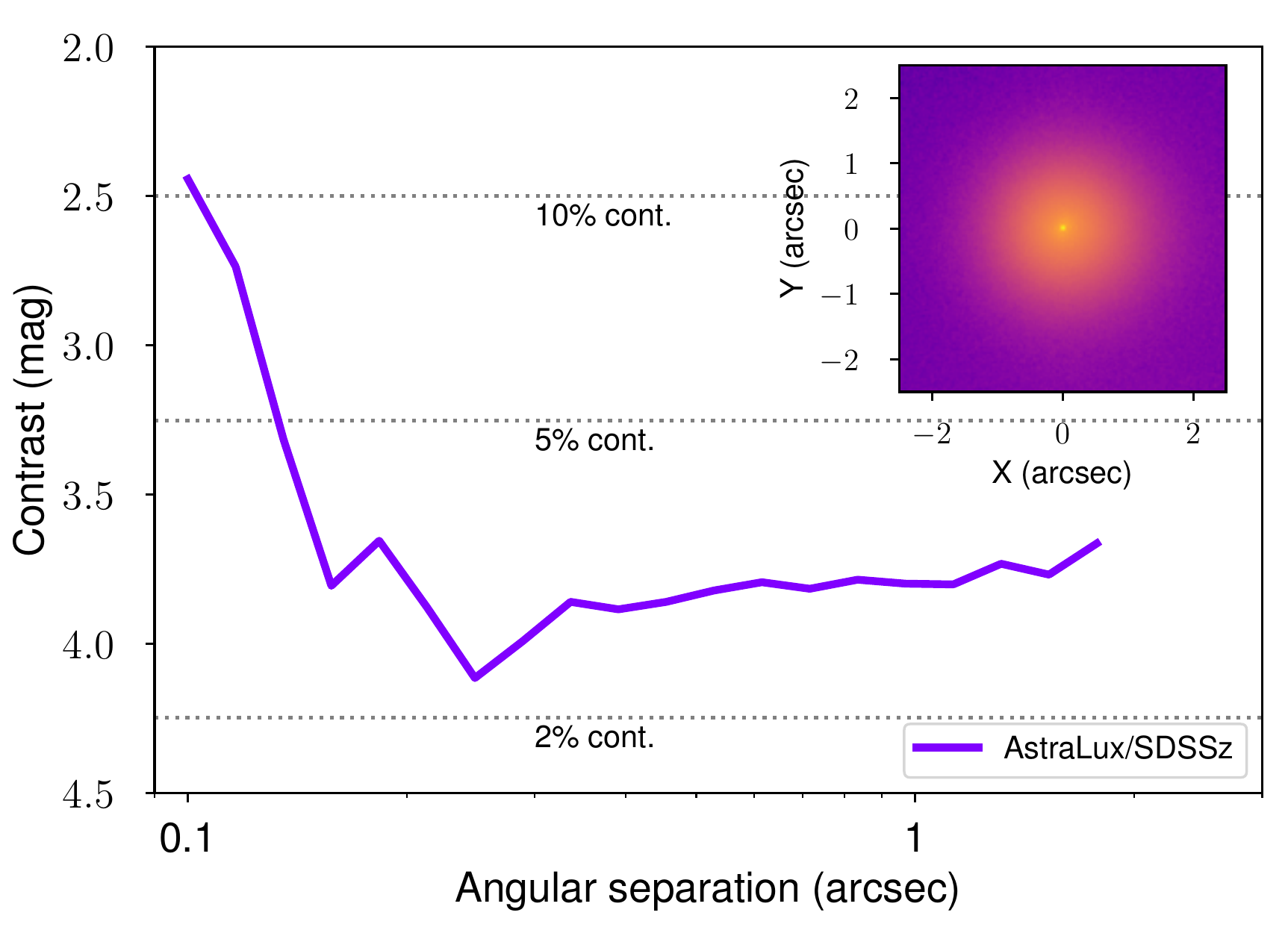}
      \caption{AstraLux $5\sigma$ sensitivity curve (main panel) of the high-spatial-resolution image obtained for LHS~1478 (inset) in the SDSSz photometric band.
      }
   \label{fig:astralux}
   \end{figure}

In order to exclude the presence of contaminants close to our target, we observed LHS~1478 with the AstraLux high-spatial-resolution camera \citep{hormuth08}, located at the 2.2\,m telescope of the Calar Alto Observatory (Almer\'{i}a, Spain). This instrument uses the lucky-imaging technique to obtain diffraction-limited images by obtaining thousands of short-exposure frames (below the atmospheric coherence time) to subsequently select those with the highest Strehl ratios \citep{strehl1902} and combine them into a final high-spatial-resolution image. We observed this target on the night of 25 February 2020 under good weather conditions with a mean seeing of 1.0\,arcsec. We obtained 41\,710 frames with 20\,ms exposure times for a total exposure of 83.4\,s in the Sloan Digital Sky Survey $z'$ filter (``SDSSz''), with a field of view windowed to $6\times6$\,arcsec. The data cube was reduced by the instrument pipeline \citep{hormuth08}, and we selected the 10\,\% of frames with the highest quality to produce the final high-resolution image. In order to obtain the sensitivity limits of the image, we used the \texttt{astrasens} package\footnote{\url{https://github.com/jlillo/astrasens}} with the procedure described in \cite{lillo-box12,lillo-box14b}. Both the 5$\sigma$ sensitivity curve and the image are shown in Fig.~\ref{fig:astralux}. We could exclude sources down to 0.2\,arcsec with a magnitude contrast of $\Delta z' < 4$\,mmag, corresponding to a maximum contamination level of 2.5\,\%.

We additionally estimated the probability of an undetected blended source in our high-spatial-resolution image, called the blended source confidence (BSC), following the steps in \cite{lillo-box14b}. In short, we used a \texttt{python} implementation of this approach (\texttt{bsc}), which uses the TRILEGAL\footnote{\url{http://stev.oapd.inaf.it/cgi-bin/trilegal}} galactic model \citep[v1.6;][]{girardi2012} to retrieve a simulated source population of the region around the corresponding target\footnote{This is done in python by using the \texttt{astrobase} implementation by \cite{astrobase}.}. 
This simulated population was used to compute the density of stars around the target position (radius $\rho=1$\,deg) and derive the probability of chance alignment at a given contrast magnitude and separation. We applied this to the position of LHS~1478 and used a maximum contrast magnitude of $\Delta m_{\rm b,max} = 6.7$\,mag in the $z'$ passband, corresponding to the maximum contrast of a blended eclipsing binary that could mimic the observed transit depth. Using our high-resolution image, we estimated the probability of an undetected blended source to be 0.2\,\%. The probability of such an undetected source being an appropriate eclipsing binary is even lower, and thus we concluded that the transit signal is not due to a blended eclipsing binary. 

\subsection{Ground-based photometry}\label{sec:photometry}

Ground-based transit photometry was obtained to confirm the TESS transit event and to refine the ephemeris of the planet. We used the TESS Transit Finder, a customized version of the {\tt Tapir} software package \citep{jensen2013}, to schedule our observations based on the preliminary ephemeris from the SPOC light curve. We obtained a total of three transit detections, which we included in our joint fit. The individual observations are described below.

   \begin{figure*}
   \centering
   \includegraphics[width=\textwidth]{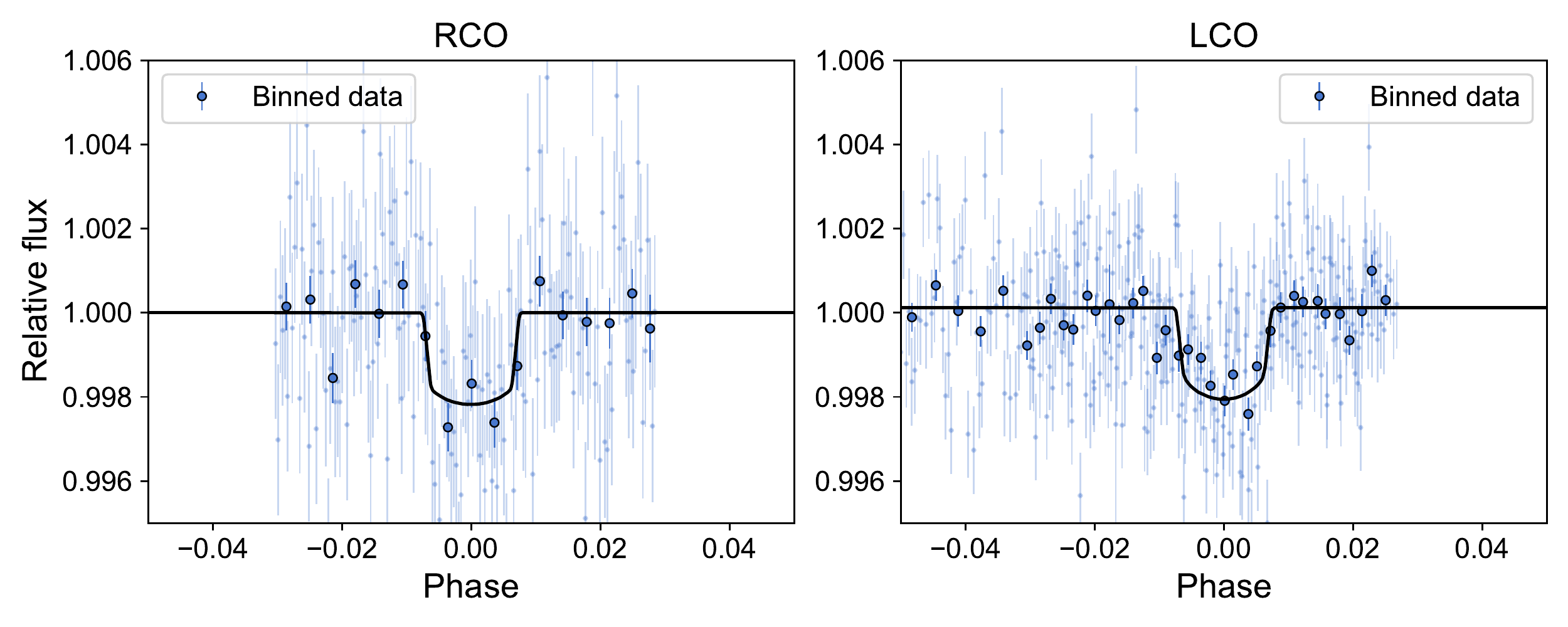}
      \caption{Phased-folded RCO ({\em left}) and LCOGT ({\em right}) data. The black line represents the best transit model from \texttt{juliet}. The small points show the raw data, while the large points show the data binned into ten-minute bins.}
         \label{fig:lcodata}
   \end{figure*}

\subsubsection{MEarth-North}

LHS~1478 was first observed by the MEarth-North telescope array on 14 January 2020. MEarth-North is located at the Fred Lawrence Whipple Observatory on Mount Hopkins, Arizona, and consists of eight 40\,cm telescopes equipped with Apogee U42 cameras and custom RG715 passbands. 
We reduced the MEarth photometry following the standard procedures outlined by \cite{irwin2007} and \citet{berta2012} and using a 6\,arcsec aperture.

LHS~1478 was observed continuously with all eight telescopes from an airmass of 1.4 to 2.0 with 14-second exposures. 
The observations were taken through intermittent clouds and produced a tentative transit detection amid large residual systematics with a photometric dispersion of 4.2\,parts-per-thousand (ppt). Although the combined light curve of the star from all eight telescopes did not yield a reliable transit detection, these data were still used to rule out nearby eclipsing binaries in 101 of 103 sources within 2.5\,arcmin and down to a differential magnitude of 8.6\,mag. 
The two uncleared sources are faint ($\Delta m$ = 3.8--4.0\,mag) and are blended with nearby sources. 
Their wide separations from LHS~1478 (1.8--2.2\,arcmin) also make them improbable sources of the TESS transit events.

\subsubsection{RCO}
The first reliable ground-based detection of a transit of LHS~1478 was obtained on 31 January 2020 using the Ritchey-Chr\'etien Optical (RCO) 40\,cm telescope located at the Grand-Pra Observatory near Valais Sion, Switzerland. 
We observed a full transit in the Sloan $i'$ passband with an exposure time of 45\,s. 
We used the {\tt AstroImageJ} software package \citep{collins2017} to perform differential photometry using 5.1\,arcsec apertures and to detrend against airmass in order to produce the light curve of LHS~1478 depicted in the left panel of Fig.~\ref{fig:lcodata}. 
With our reduction, we achieved a photometric precision of 1.1\,ppt in three-minute bins. 
We detected the transit of LHS~1478 with a mid-transit time of $\approx 8$\,min late relative to the SPOC prediction (the transit epoch was derived at the time using TESS data from sectors 18 and 19). 

\subsubsection{LCOGT}
We observed two additional full transits of LHS~1478 on 27 August 2020 and 7 October 2020 from the Las Cumbres Observatory Global Telescope \citep[LCOGT;][]{brown2013} 1.0\,m network node at McDonald Observatory. The images were calibrated by the standard LCOGT {\tt Banzai} pipeline \citep{mccully2018}.

Both light curves were obtained in the Pan-STARRS $z$-short passband with exposure times of 40\,s and 50\,s in the August and October runs, respectively. We used {\tt AstroImageJ} to perform differential photometry using 5.8\,arcsec apertures and to detrend against airmass. The resulting combined and phase-folded light curves are included in the right panel of Fig.~\ref{fig:lcodata}. With our reduction, we achieve photometric precisions of 0.7\,ppt and 0.5\,ppt in three-minute bins, which resulted in a high signal-to-noise ratio (S/N) transit detection in each LCOGT light curve. Both LCOGT transit detections were at the time predicted by the refined ephemeris of LHS~1478 from the RCO detection.

\subsection{Spectroscopy}

\subsubsection{CARMENES}\label{sec:carmenesobs}

We obtained a total of 57 spectra of LHS~1478 with CARMENES between 28 January 2020 and 16 October 2020, all taken with exposure times of 1800\,s. 
The visual channel (VIS) of CARMENES has a resolving power $R$ = 94\,600 and covers the spectral range from 520\,nm to 960\,nm. 
Simultaneous wavelength calibration is performed with Fabry-P\'erot etalon exposures, which are also used to track the instrument drift at night. The obtained spectra have a median S/N of $\sim$ 48 per pixel at 746\,nm. The spectra were reduced with \texttt{caracal} \citep{Zechmeister2014,Caballero2016b}, and the corresponding RVs were extracted using \texttt{serval} \citep{Zechmeister2018} along with selected activity indicators (Sect.~\ref{sec:stellar_params}). 
The RVs were corrected for barycentric motion, instrumental drift, secular acceleration, and nightly zero points \citep{tal-or2019, trifonov2020}. The root-mean-square uncertainty (rms) and median uncertainty ($\hat{\sigma}$) of the VIS RVs were 4.02\,m\,s$^{-1}$ and 2.49\,m\,s$^{-1}$, respectively, and are shown in Fig.~\ref{fig:joint_fit} and Table~\ref{tab:carmenes_data}.
A generalized Lomb-Scargle (GLS) periodogram \citep{Zechmeister2009} of the RV data shows a period at 1.949\,d, which is consistent with the transit signal, with a theoretically computed false-alarm probability (FAP) of less than 1\,\% (Fig.~\ref{fig:gls_activity}). 

The RVs measured with the CARMENES near-infrared (NIR) channel have rms and median uncertainties of 19.26\,m\,s$^{-1}$ and 9.97\,m\,s$^{-1}$, respectively. 
We excluded the NIR RVs from the analysis because their scatter is higher than the expected planetary amplitude.

\subsubsection{IRD spectroscopy}\label{sec:irdobs}

We observed LHS~1478 using the InfraRed Doppler (IRD) instrument at the Subaru 8.2\,m telescope \citep{2012SPIE.8446E..1TT, 2018SPIE10702E..11K} between September and December 2020. A total of 13 spectra were obtained, with an integration time of 900\,s. The IRD instrument is a fiber-fed spectrograph that covers the spectral range from 930\,nm to 1740\,nm at a resolving power of $R \sim$ 70\,000. Simultaneous wavelength calibration is performed by injecting light from a laser-frequency comb into the second fiber. 
The raw data were reduced into one-dimensional spectra using the {\tt IRAF} software, as well as with a custom code \citep{2018SPIE10702E..60K, 2020PASJ...72...93H} to suppress the bias and correlated noise of the detectors. 
The typical S/N of the reduced spectra ranged from 60 to 95 per pixel at 1000\,nm.  

Following \citet{2020PASJ...72...93H}, we analyzed the observed spectra to extract precise RVs. In doing so, we combined multiple frames to obtain a high S/N template for the RV analysis, after removing the telluric lines and the instrumental profile of the spectrograph. The relative RV was then measured with respect to this template for each spectrum. The resulting relative RVs after the correction for the barycentric motion of Earth are listed in Table \ref{tab:ird_data} and shown in Fig.~\ref{fig:joint_fit}. The internal RV error was typically 3--6\,m\,s$^{-1}$ for each frame.

\section{Analysis}

\begingroup
\renewcommand{\arraystretch}{1.2}
\begin{table}
\caption{Stellar parameters of LHS~1478.}\label{tab:stellar_parameters}
\centering
\begin{tabular}{lcr}
\hline\hline
Parameter & Value & Reference\\
\hline
Name & LHS~1478 & Luy79\\
Karmn & J02573+765 & Cab16 \\
TOI & 1640 & ExoFOP-TESS\\
TIC & 396562848 & Sta18\\
\noalign{\smallskip}

$\alpha$ (J2000) & 02 57 17.51 & \textit{Gaia} EDR3\\
$\delta$ (J2000) & +76 33 13.0 & \textit{Gaia} EDR3\\
Sp. type & m3\,V\tablefootmark{a} & This work \\
$J$ [mag] & $9.615\pm0.026$ & Skr06\\ 
$G$ [mag] & $12.2481\pm0.0028$ & \textit{Gaia} EDR3\\
$T$ [mag] & $11.0548\pm0.0073$ & Sta19\\
\noalign{\smallskip}

$\varpi$ [mas] & $54.904\pm0.018$ & \textit{Gaia} EDR3\\
$d$ [pc] & $18.214\pm0.006$ & \textit{Gaia} EDR3\\
\noalign{\smallskip}

$T_{\text{eff}}$ [K] & $3381\pm 54$ & This work \\
$\log{g}$ [cgs] & $4.87 \pm 0.06$ & This work \\
{[Fe/H]} [dex] & $-0.13 \pm 0.19$ & This work \\
\noalign{\smallskip}

$M_{\star}$ [$M_{\odot}$] & $0.236 \pm 0.012$ & This work \\
$R_{\star}$ [$R_{\odot}$] & $0.246 \pm 0.008$ & This work \\
$\rho_{\star}$ [g cm$^{-3}$]\tablefootmark{b} & $22.2 \pm 2.5$ & This work\\
$L_{\star}$ [$10^{-4}\, L_{\odot}$] &  $71.5 \pm 1.2$ & Cif20\\ 
pEW'(H$\alpha$) [\textup{\AA}] & $+0.040 \pm 0.026$ & This work \\
$v\sin i$ [km\,s$^{-1}$] & $< 2$ & Mar21\\
$P_{\text{rot}}$ [d] & [6.4]\tablefootmark{c} & New16\\

\noalign{\smallskip}
\hline

\end{tabular}
\tablebib{
Luy79: \citet{Luyten1979}; 
Sta18: \citet{Stassun2018}; 
Cab16: \citet{Caballero2016a}; 
Skr06: \citet{Skrutskie2006}; 
\textit{Gaia} EDR3: \cite{GaiaEDR3}; 
Sta19: 
\citet{Stassun2019}; 
Cif20: \citet{Cifuentes2020}; 
Mar21: Marfil et al. (in prep); 
New16: \citet{Newton2016}.
}
\tablefoot{
\tablefoottext{a}{Spectral type estimated from photometry and parallax (with a lowercase ``m''), as by \citet{Cifuentes2020}.}
\tablefoottext{b}{Derived from $M_{\star}$ and $R_{\star}$.}
\tablefoottext{c}{Flagged as a non-detection or an undetermined detection by \citet{Newton2016}.}
}
\end{table}
\endgroup

\begingroup
\renewcommand{\arraystretch}{1.3}
\begin{table}
\caption{Posterior distributions from the joint fit. The uncertainties represent the 68\% CI of the obtained distributions.}\label{tab:planet_parameters}
\centering
\begin{tabular}{lc}
\hline\hline
Parameter & Value\\
\hline
\noalign{\smallskip}
\multicolumn{2}{c}{\em Stellar density}\\
\noalign{\smallskip}
$\rho_{\star}$ [g\,cm$^{-3}$]\tablefootmark{a} & $22.2^{+2.7}_{-2.3}$ \\
\noalign{\smallskip}
\multicolumn{2}{c}{\em Orbital parameters}\\
\noalign{\smallskip}
$P$ [d] & $1.9495378^{+0.0000040}_{-0.0000041}$ \\
$T_0$ [BJD] & $2458786.75425^{+0.00042}_{-0.00042}$ \\
$r_1$\tablefootmark{b} & $0.8111^{+0.0079}_{-0.0088}$ \\
$r_2$\tablefootmark{b} & $0.0462^{+0.0011}_{-0.0010}$ \\
$p = R_p/R_{\star}$ & $0.0462^{+0.0011}_{-0.0010}$ \\
$b = (a/R_{\star})\cos i$ & $0.717^{+0.012}_{-0.013}$ \\
$i$ [deg] & $87.452^{+0.052}_{-0.048}$ \\
$a/R_{\star}$ & $16.119^{+0.088}_{-0.094}$ \\
$K$ [m s$^{-1}$] & $3.13^{+0.23}_{-0.25}$ \\
$t_T$ [h] & $0.705^{+0.011}_{-0.011}$ \\

\noalign{\smallskip}
\multicolumn{2}{c}{\em Derived planetary parameters}\\
\noalign{\smallskip}

$M_p$ [$M_{\oplus}$] & $2.33^{+0.20}_{-0.20}$ \\
$R_p$ [$R_{\oplus}$] & $1.242^{+0.051}_{-0.049}$ \\
$\rho_p$ [g cm$^{-3}$] & $6.67^{+1.03}_{-0.89}$ \\
$a$ [au] & $0.01848^{+0.00061}_{-0.00063}$ \\
$T_{\text{eq}}$ [K]\tablefootmark{c} & $595^{+10}_{-10}$ \\

\noalign{\smallskip}
\hline
\end{tabular}
\tablefoot{
\tablefoottext{a}{Derived from light curve fitting using the relations from \citet{Seager2003}.}
\tablefoottext{b}{Parameterization from \citet{Espinoza2018} for $p$ and $b$.}
\tablefoottext{c}{Assuming zero Bond albedo.}
}
\end{table}
\endgroup

\subsection{Stellar parameters}\label{sec:stellar_params}

The stellar parameters for this target were estimated using the CARMENES VIS stacked stellar template produced by \texttt{serval}. The 
$T_{\text{eff}}$, $\log g$, and iron abundance [Fe/H] were determined through spectral fitting with a grid of PHOENIX-SESAM models following \citet{Passegger2019}, using the upper limit $v \sin i = 2.0$\,km\,s$^{-1}$ from \citet{Reiners2018} and Marfil et al (in prep.), who did not detect any rotational velocity. 
The luminosity, $L_{\star}$, was estimated by integrating the spectral energy distribution with the photometric data used by \citet{Cifuentes2020}. 
The stellar radius, $R_{\star}$, was determined via the Stefan–Boltzmann law, and the stellar mass $M_{\star}$ via the mass-radius relation from \citet{Schweitzer2019}.
We estimated the overall activity level of the star from the pseudo-equivalent width of the H$\alpha$ line after the subtraction of an inactive stellar template, pEW'(H$\alpha$), following \citet{schofer2019}. We obtained a value of $+0.040 \pm 0.026$\,\textup{\AA}, which indicates that LHS~1478 is a fairly inactive star \citep{jeffers2018,schofer2019}.

The star has a rotational period of 6.4\,d, as listed by \citet{Newton2016}, but the detection is deemed inconclusive. We looked at the SAP (Simple Aperture Photometry; \citealt{morris2020}) and PDC data from TESS and, after masking the transits and performing a Lomb-Scargle periodogram, we could not detect any signal with a period of 6.4\,d. 
Therefore, we agreed with \citet{Newton2016} in that the signal is a non-detection and may not correspond to the real rotational period for this star. 

Time series of three activity indicators were extracted from the spectra with \texttt{serval}: the chromatic index (CRX), differential line width (dLW), and H$\alpha$ \citep{Zechmeister2018}. 
We found no signs of periodic variations similar to the planet orbital period or the putative stellar rotational period of 6.4\,d (Fig.~\ref{fig:gls_activity}). 
The only exception is for the H$\alpha$ data, where there is a signal at $\approx$ 6.8\,d with an FAP = 5\,\%. 
This signal is not seen in any of the other indices or data sets, and, therefore, we could not provide any insight into its origin or relevance. 
The activity data are not correlated with the RV data either, with a correlation coefficient of $| r | < 0.3$.
A summary of the stellar properties and relevant photometric and astrometric data from \textit{Gaia} EDR3 is shown in Table~\ref{tab:stellar_parameters}.

\subsection{Joint fit}\label{sec:joint-fit}

   \begin{figure}
   \centering
   \includegraphics[width=\hsize]{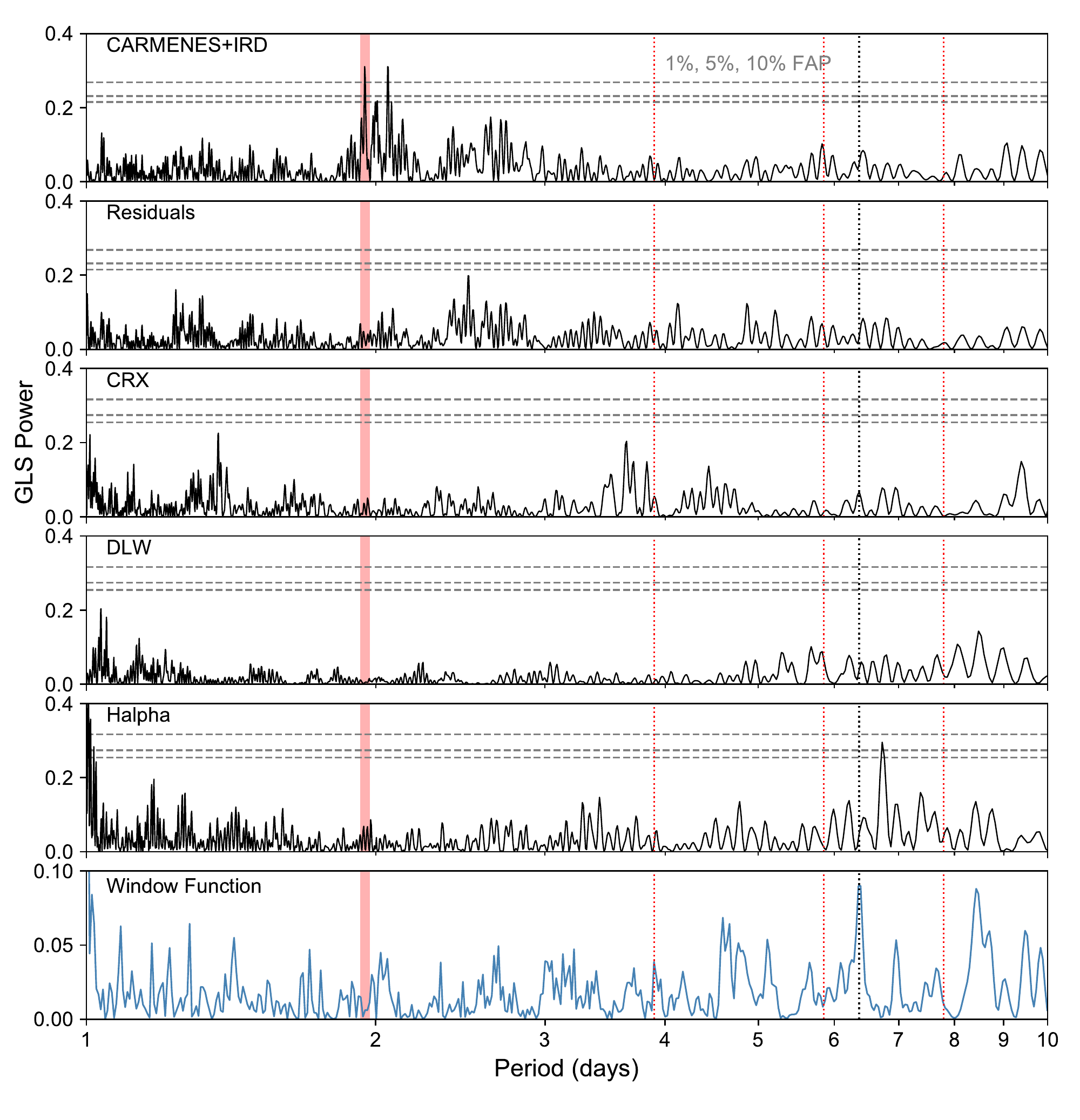}
      \caption{GLS periodograms for the combined RVs, RV residuals from the fit, the activity indices CRX, dLW, and H$\alpha$, and the window function.
      The dotted horizontal lines represent the 1\,\%, 5\,\%, and 10\,\% FAP. 
      The red shaded area is the position of the orbital period of the planet, and the vertical dotted red lines are aliases of that period. The vertical dotted black  line represents the 6.4\,d stellar rotation period by \citet{Newton2016}.
      }
         \label{fig:gls_activity}
   \end{figure}
   
We used \texttt{juliet} \citep{juliet} to perform a joint fit of the TESS, RCO, LCOGT, CARMENES, and IRD data.
We used the efficient, uninformative sampling scheme from \citet{Kipping2013} together with a quadratic law to parameterize the limb darkening in the TESS data. For the RCO and LCOGT data we used a linear limb-darkening law. We also followed the parameterization presented in \citet{Espinoza2018}, with parameters $r_1$ and $r_2$, to fit for $p$ and $b$, the planet-to-star radius ratio and impact parameter, respectively.
The \tess data were fitted separately by sector, each with its own relative flux offset and jitter, but we imposed identical limb-darkening coefficients for the transit curves from all sectors. 
We fitted the LCOGT data separately for each observing night, but, as with the TESS data, we imposed an identical limb-darkening coefficient for the two nights.
The detrending of the TESS data was done by incorporating a Gaussian process (GP) model.
We used the \texttt{celerite} M\'atern kernel with a hyperparameter amplitude and the GP timescale given by $\sigma_{\text{GP}}$ and $\rho_{\text{GP}}$, respectively \citep{celerite}. 
The same GP model was used for each \tess sector.
An initial search for the optimal parameters of this system was done with \texttt{exostriker} \citep{exostriker} using all data sets, and the results were used to build the priors for \texttt{juliet}, which are shown in Table~\ref{tab:juliet_priors}. The obtained posterior probabilities are listed in Tables~\ref{tab:planet_parameters} and~\ref{tab:inst_parameters}.

We built our full model, shown in Figs.~\ref{fig:joint_fit} and \ref{fig:lcodata}, from the 68\,\% confidence interval (CI) of the posterior distribution for each parameter. 
We found that the signal observed in both the photometric and RV data is consistent with a $2.33^{+0.20}_{-0.20}\,M_{\oplus}$ and $1.242^{+0.051}_{-0.049}\, R_{\oplus}$ planet, with a density of $6.67^{+1.03}_{-0.89}$\,g\,cm$^{-3}$, orbiting the star with a period of $1.9495378^{+0.0000040}_{-0.0000041}$\,d.

Since there is a brighter star ($G \approx 10.30$\,mag) about 120\,arcsec north of LHS\,1478, we checked for possible contamination in the TESS light curve. We analyzed the transit data from TESS and LCOGT, using the GP-detrended data, leaving the dilution factor as a free parameter for the TESS data, and fixing the limb-darkening coefficients to the values interpolated from limb darkening tables \citep{Claret2013,Claret2017}. The parameters from Table\,\ref{tab:planet_parameters} were recovered very well, and the best fit dilution factor is below 1\,\% (and below 13\,\% at 95\,\% confidence). We conclude that the TESS light curve is therefore not contaminated.

A TLS search in the light curve residuals from the TESS data showed two significant periods at 13.9\,d and 27.8\,d, but they were due to the sampling of the data (Fig.~\ref{fig:TLS_residuals}). 
A GLS periodogram of the residuals of the RV data did not show any significant signals either (Fig.~\ref{fig:gls_activity}).
The obtained jitter terms for the RVs are within the expected instrumental jitter intervals for each instrument:$\sim$ 1.2\,m\,s$^{-1}$ for CARMENES VIS (\citealt{Bauer2020}) and $\sim$ 2--3\,m\,s$^{-1}$ for IRD (\citealt{2020PASJ...72...93H}).

\section{Discussion and characterization prospects}\label{sec:discussion}
 
   \begin{figure}
   \centering
   \includegraphics[width=\hsize]{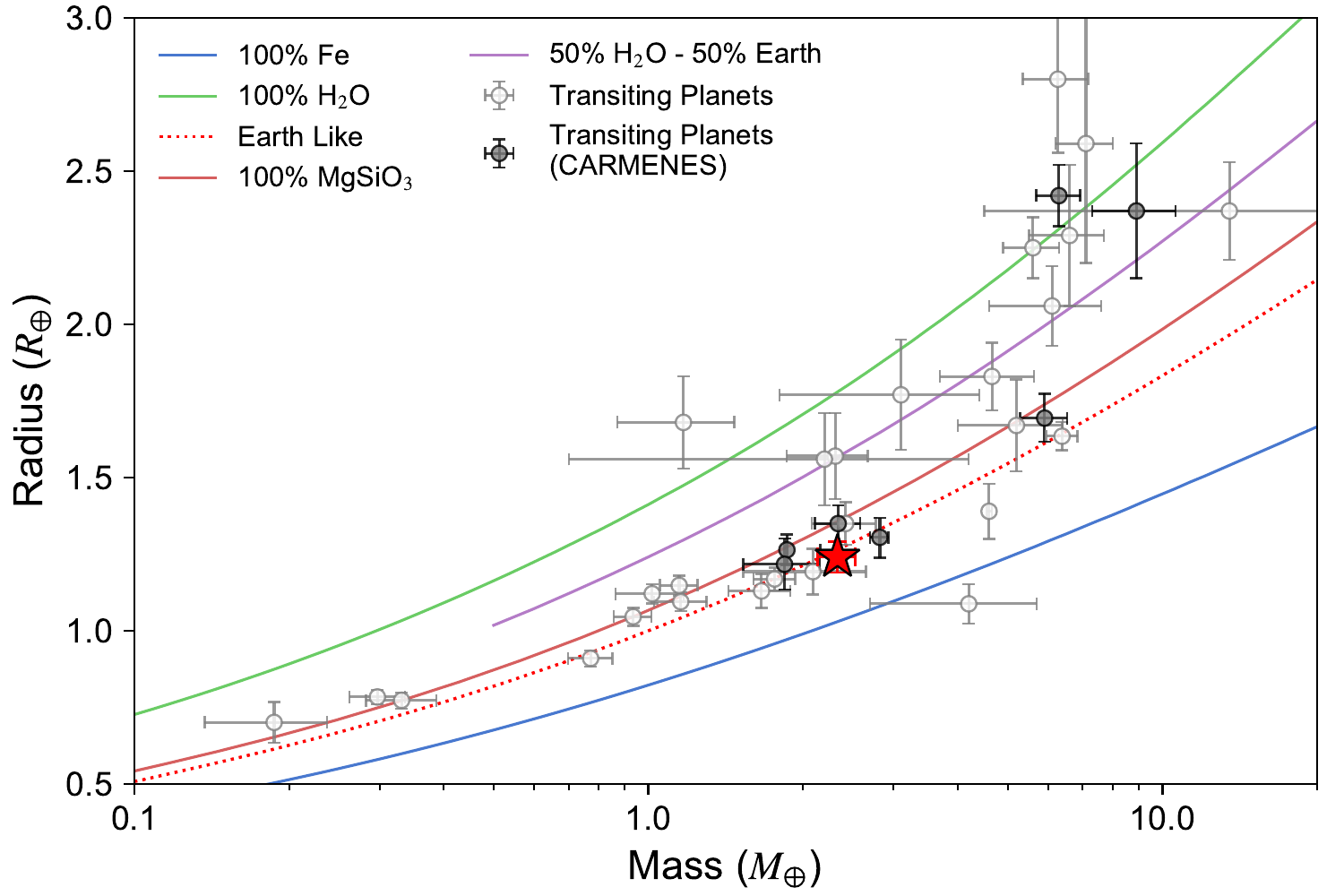}
      \caption{Mass-radius diagram for LHS~1478~b (red star). The colored lines represent the composition models from \citet{Zeng2016, Zeng2019}. Transiting planets around M dwarfs with mass and radius measurements are shown in black circles (detected with CARMENES) and gray circles.}
      \label{fig:composition}
   \end{figure}
   
Given the measured properties, we made a first exploratory guess of the planet composition. We compared its mass and radius with the models from \citet{Zeng2016, Zeng2019}, shown in Fig.~\ref{fig:composition}, plus other transiting planets around M dwarfs from the literature\footnote{Data on transiting M dwarf planets at \url{https://carmenes.caha.es/ext/tmp/}.}. We found that LHS~1478~b is compatible with a bulk composition of $\sim$30\,\% Fe plus 70\,\% MgSiO$_3$.\ This makes its composition comparable to Earth's, thus strongly supporting the notion that it is indeed a rocky world. 


   \begin{figure}
   \centering
   \includegraphics[width=\hsize]{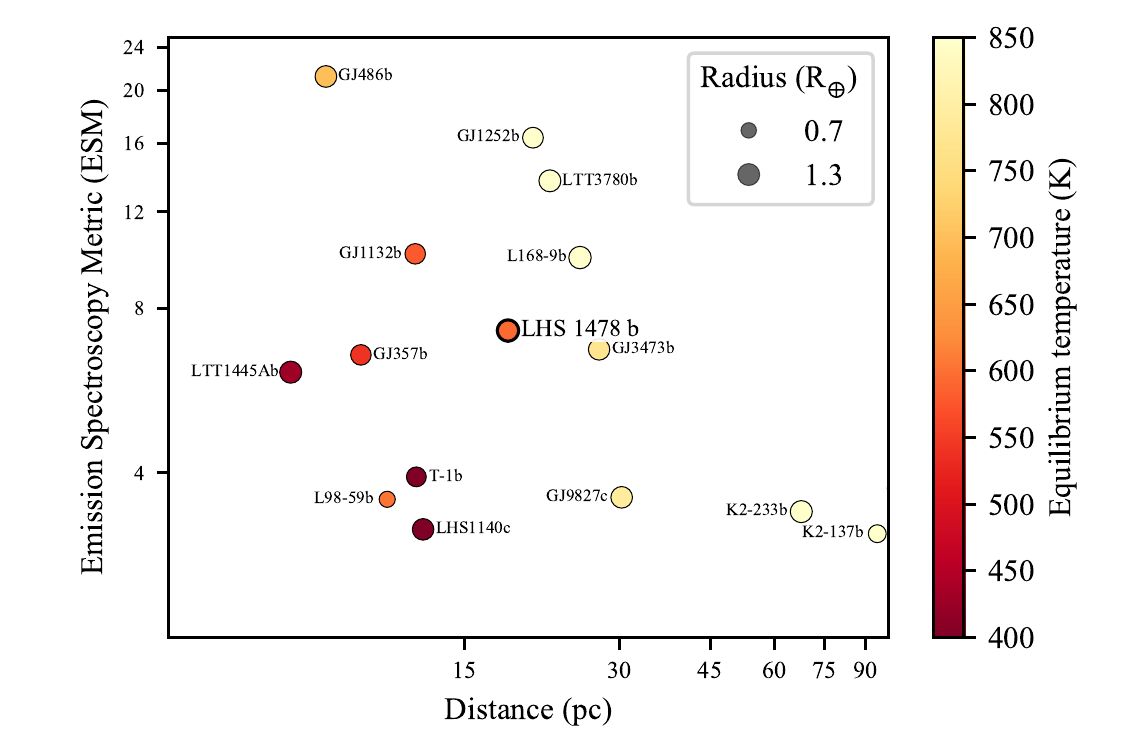}
   \includegraphics[width=\hsize]{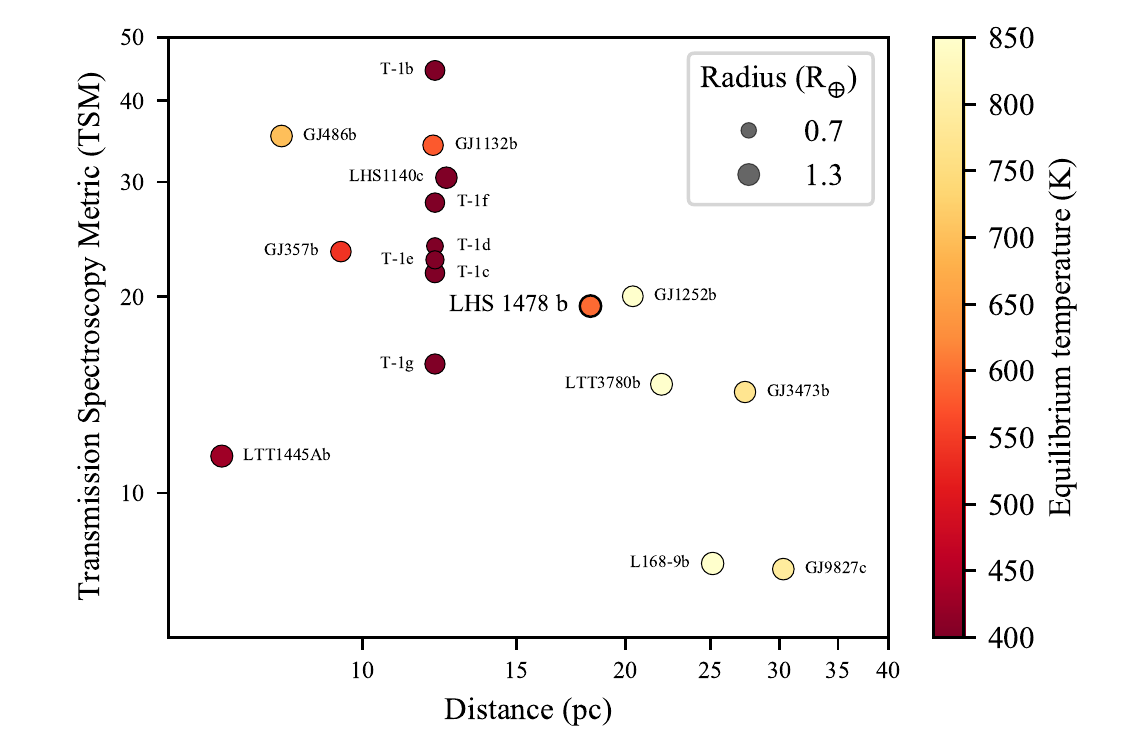}
      \caption{ESM ({\em top panel}) and TSM ({\em bottom panel}) as a function of distance from the Sun for exoplanets with radii of less than 1.5\,$R_\oplus$ and mass determinations by RVs or TTVs. LHS~1478~b is labeled and marked with a thicker black borderline in both panels.}
         \label{fig:TSM_ESM}
   \end{figure}
   
   \begin{figure*}
   \centering
   \includegraphics[width=\hsize]{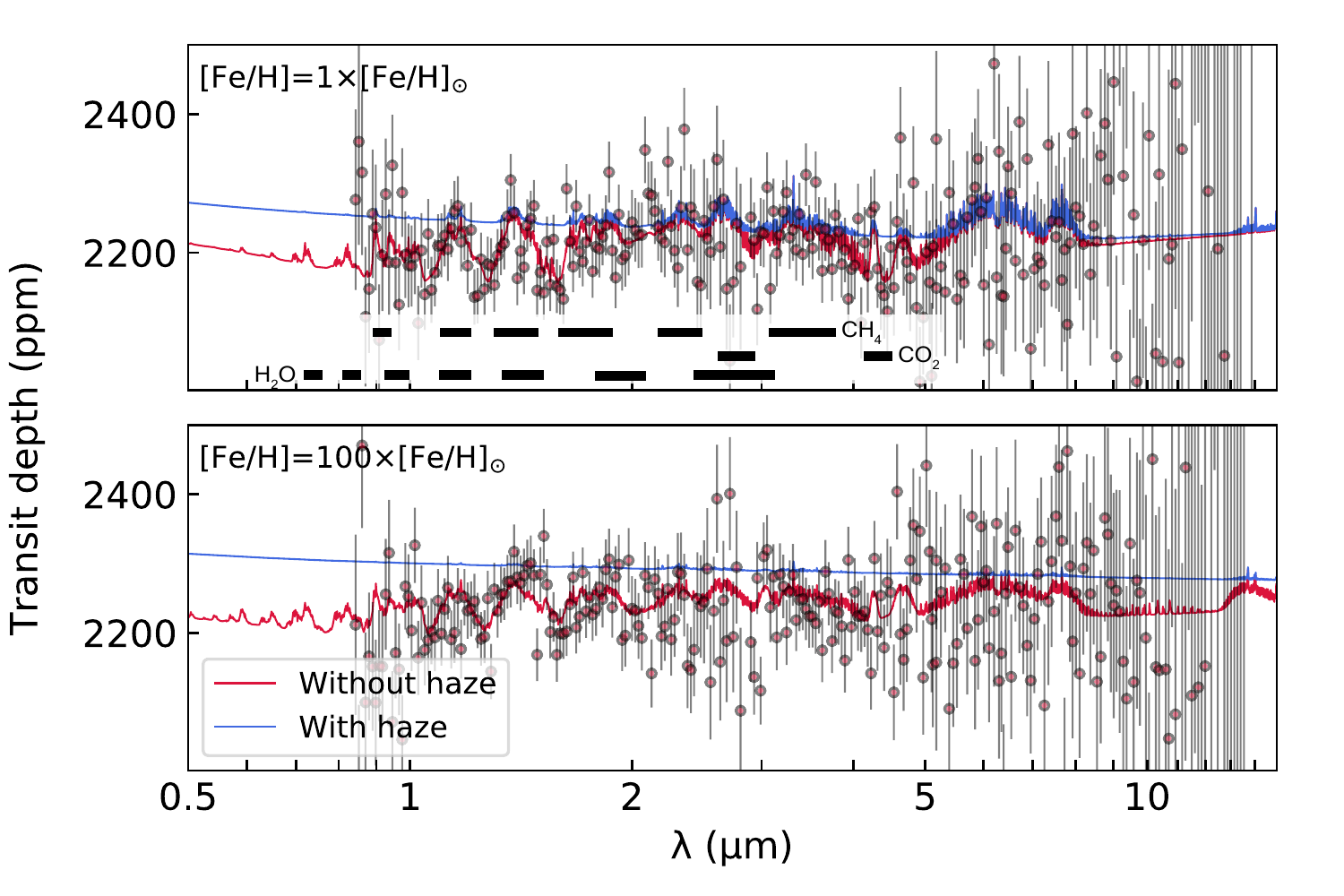}
      \caption{ Synthetic transmission atmospheric spectra of LHS~1478~b with haze opacity (solid blue lines) and without haze opacity (solid red lines). Simulated observations and estimated uncertainties are shown for the {\em JWST} NIRISS/SOSS (0.6--2.8\,$\mu$m), NIRSpec/G395M (2.87--5.10\,$\mu$m), and MIRI/LRS (5--12\,$\mu$m) configurations, assuming two transits and binned for $R$ = 50 (filled circles with error bars).
      {\em Top}: Fiducial models with solar abundance and the location of the strongest molecular features of H$_2$O, CH$_4$, and CO$_2$. 
      {\em Bottom}: Metallicity enhanced by a factor of 100.}
         \label{fig:Synthetic_spec_TOI1640}
   \end{figure*}
   
Atmospheric characterization of rocky planets with (some) properties similar to Earth's is one of the pivotal developments expected in the forthcoming years thanks to the deployment of new ground- and space-based facilities. 
The potential of a target for detailed characterization is a complicated function of the planet and host star properties, as well as the instrument to be used. In this sense, not all transiting planets that have interesting properties are equally suitable for actual characterization.

As a first approximation of the suitability of LHS~1478~b, we calculated the spectroscopic metrics from \citet{Kempton2018}, which were developed to rank the best TESS targets for the instrumentation aboard the {\em James Webb Space Telescope} ({\em JWST}). We estimated the emission spectroscopy metric (ESM) and the transmission spectroscopy metric (TSM) to be 7.28 and 19.35, respectively. 
The upper panel of Fig.~\ref{fig:TSM_ESM} shows the ESMs of rocky exoplanets with measured masses, either through RVs or TTVs, and puts LHS~1478~b in that context. 
The ESM of 7.28 is slightly lower than the 7.5 threshold set by \citet{Kempton2018}, but it is very close to that of Gl~1132~b, which was considered by the authors as a benchmark rocky planet for emission spectroscopy (after performing a recomputation of the ESM for Gl~1132~b, we found a value of 10.06, which is less than half of the ESM value of \object{GJ~486}\,b, recently discovered by \citealt{Trifonov2021}).
The lower panel illustrates TSM values. 
An acceptable TSM value for this class of planet is around 12 or higher \citep[Table~1 in][]{Kempton2018}, which implies that LHS~1478~b is likely also an appropriate candidate for atmospheric characterization through transmission spectroscopy.

As a second refinement to its suitability for characterization, we assessed the potential chemical species that could be detected in its atmosphere using near-future instrumentation. Molecular features such as water, carbon dioxide, or methane should typically be observable on these kinds of planets if they maintain a substantial atmosphere \citep{molaverdikhani2019coldb,molaverdikhani2019colda}, but the detectability of such features could also be obscured by the presence of clouds \citep{molaverdikhani2020role}. 
In order to quantitatively assess the potential of the atmospheric characterization observations of LHS~1478~b with the {\em JWST}, we calculated a few atmospheric models and their compositions using the photochemical code {\tt ChemKM} \citep{molaverdikhani2019coldb} and their corresponding spectra using {\tt petitRADTRANS} \citep{molliere2019petitradtrans}. We assumed a stellar radiation environment similar to that measured for the well-studied star \object{GJ~667C} ($\rm T_{eff} \approx$~3327\,K). For the model of the planetary atmospheric structure, we adapted a Venus-like temperature profile to make it consistent with the derived brium temperature of the planet (see Table~\ref{tab:planet_parameters}). 
Our photochemical models included 135 species and 788 reactions.

Figure~\ref{fig:Synthetic_spec_TOI1640} shows simulated realizations of the transmission spectrum of LHS~1478~b, assuming solar metallicities (top panel) and 100-times-enhanced metallicities (bottom panel). 
The dominant spectral features come from water and methane, as expected, with a less extended CO$_2$ feature at around 4.5\,$\mu$m. 
The amplitudes of these features are between 50 and 100\,ppm, which should be measurable given that the star is rather bright. This was verified using {\tt PandExo} \citep{batalha2017pandexo} simulations of the NIRISS/SOSS, NIRSpec/G395M, and MIRI/LRS instruments and modes on board {\em JWST}.

Hydrogen-dominated atmospheres with temperatures below 900\,K are expected to have significant photochemically produced hazes \citep[e.g.,][]{he2018photochemical,gao2020aerosol}. If such a scenario applied to LHS~1478~b, it would result in the obscuration of spectral features, particularly in the case of an enhanced metallicity (see the bottom panel of Fig.~\ref{fig:Synthetic_spec_TOI1640}). A relatively flat transmission spectrum thus might be indicative of a metallicity-enhanced atmosphere covered by haze, which might be indistinguishable from the planet having no atmosphere at all. 
If a (nearly) flat spectrum were observed, ground-based high-resolution spectroscopy could then be used to detect unobscured molecular features and break the associated degeneracies (see \citealt{Brogi2017} and \citealt{Gandhi2020} for more details on the possible features that could be detected).

In short, the simulations described in this section indicate that LHS~1478~b, along with GJ~357~b \citep{luque:2019}, GJ~1132~b \citep{berta2015rocky}, and GJ~486~b \citep{Trifonov2021}, belongs to the small family of planets where we can realistically expect to obtain meaningful measurements and constraints with next-generation space telescopes such as {\em JWST}.

\section{Conclusions}\label{sec:conclusions}

In this paper we present TESS and ground-based photometric observations, together with CARMENES and IRD Doppler spectroscopy, of the star LHS~1478. 
We determine LHS~1478~b to have a mass of $2.33^{+0.20}_{-0.19}\,M_{\oplus}$, a radius of $1.242^{+0.050}_{-0.049}\, R_{\oplus}$, and a bulk density of $6.67^{+1.03}_{-0.89}$\,g\,cm$^{-3}$, which is consistent with an Earth-like composition. The star is remarkably inactive, and we see no signs of additional planetary signals in the photometry or the RVs, which thus greatly simplifies the analysis.

The equilibrium temperature of this planet places it in a recently found group of warm rocky planets, together with GJ~357~b, GJ~1132~b, and GJ~486~b, which are ideal for atmospheric studies. The fact that the planetary signal is very isolated in both photometry and RV, with no stellar activity or additional companions contaminating it, together with the proximity of the system to the Sun and the relative brightness of the star make this an ideal target for near-future transit observations with {\em JWST} and ground-based high-resolution spectrometers.

\begin{acknowledgements}
CARMENES is an instrument at the Centro Astron\'omico Hispano-Alem\'an (CAHA) at Calar Alto (Almer\'{\i}a, Spain), operated jointly by the Junta de Andaluc\'ia and the Instituto de Astrof\'isica de Andaluc\'ia (CSIC).
  
  CARMENES was funded by the Max-Planck-Gesellschaft (MPG), 
  the Consejo Superior de Investigaciones Cient\'{\i}ficas (CSIC),
  the Ministerio de Econom\'ia y Competitividad (MINECO) and the European Regional Development Fund (ERDF) through projects FICTS-2011-02, ICTS-2017-07-CAHA-4, and CAHA16-CE-3978, 
  and the members of the CARMENES Consortium 
  (Max-Planck-Institut f\"ur Astronomie,
  Instituto de Astrof\'{\i}sica de Andaluc\'{\i}a,
  Landessternwarte K\"onigstuhl,
  Institut de Ci\`encies de l'Espai,
  Institut f\"ur Astrophysik G\"ottingen,
  Universidad Complutense de Madrid,
  Th\"uringer Landessternwarte Tautenburg,
  Instituto de Astrof\'{\i}sica de Canarias,
  Hamburger Sternwarte,
  Centro de Astrobiolog\'{\i}a and
  Centro Astron\'omico Hispano-Alem\'an), 
  with additional contributions by the MINECO, 
  the Deutsche Forschungsgemeinschaft through the Major Research Instrumentation Programme and Research Unit FOR2544 ``Blue Planets around Red Stars'', 
  the Klaus Tschira Stiftung, 
  the states of Baden-W\"urttemberg and Niedersachsen, 
  and by the Junta de Andaluc\'{\i}a.

  This paper included data collected by the TESS mission. Funding for the TESS mission is provided by the NASA's Science Mission Directorate.
  Resources supporting this work were provided by the NASA High-End Computing (HEC) Program through the NASA Advanced Supercomputing (NAS) Division at Ames Research Center for the production of the SPOC data products.
  
  This work made use of observations from the LCOGT network.
  LCOGT telescope time was granted by NOIRLab through the Mid-Scale Innovations Program (MSIP), which is funded by the National Science Foundation.
  
  We acknowledge financial support from
  STFC grants ST/P000592/1 and ST/T000341/1,
  NASA grant NNX17AG24G,
  the Agencia Estatal de Investigaci\'on of the Ministerio de Ciencia, Innovaci\'on y Universidades and the ERDF through projects 
  PID2019-109522GB-C5[1:4]/AEI/10.13039/501100011033,   
  PGC2018-098153-B-C3[1,3]  
and the Centre of Excellence ``Severo Ochoa'' and ``Mar\'ia de Maeztu'' awards to the Instituto de Astrof\'isica de Canarias (SEV-2015-0548), Instituto de Astrof\'isica de Andaluc\'ia (SEV-2017-0709), and Centro de Astrobiolog\'ia (MDM-2017-0737), 
    the Generalitat de Catalunya/CERCA programme, 
    Grant-in-Aid for JSPS Fellows Grant (JP20J21872), 
    JSPS KAKENHI Grant (22000005, JP15H02063,JP18H01265, JP18H05439, JP18H05442, JP19K14783), 
    JST PRESTO Grant (JPMJPR1775),
    and University Research Support Grant from the National Astronomical Observatory of Japan.

\end{acknowledgements}

\bibliographystyle{aa} 
\bibliography{toi1640}

\begin{appendix}
\section{Additional figures and tables}

   \begin{figure}[]
   \centering
   \includegraphics[width=\hsize]{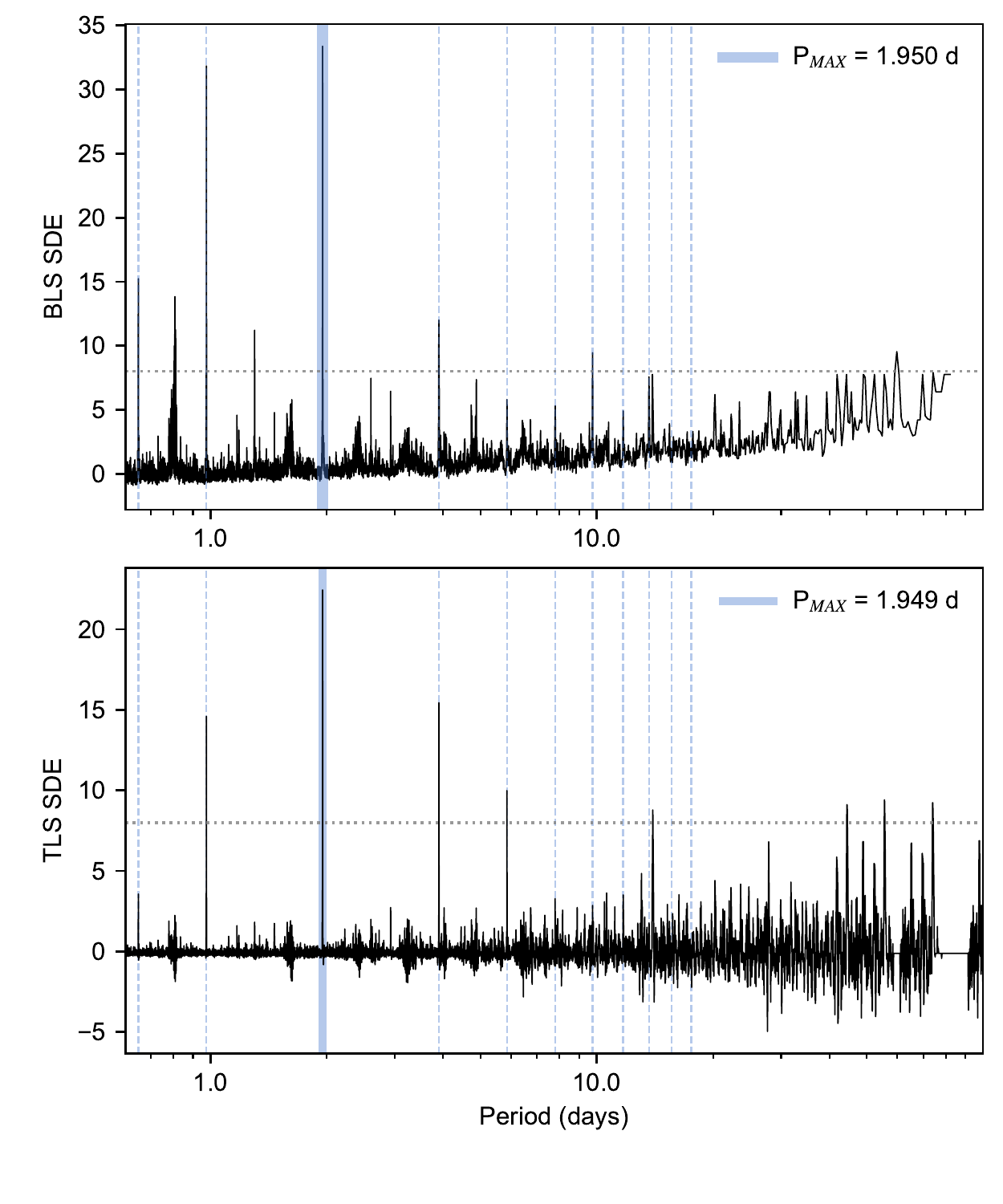}
      \caption{BLS ({\em top panel}) and TLS ({\em bottom panel}) SDE for the \tess light curve. The vertical blue lines correspond to aliases of the maximum period, highlighted by the blue shaded region. The dashed horizontal line at SDE = 8 in both panels corresponds to the threshold for transit detection from \citet{aigrain2016}.}
         \label{fig:BLS_TLS}
   \end{figure}

\begin{figure*}
    \centering
    \includegraphics[width=\textwidth]{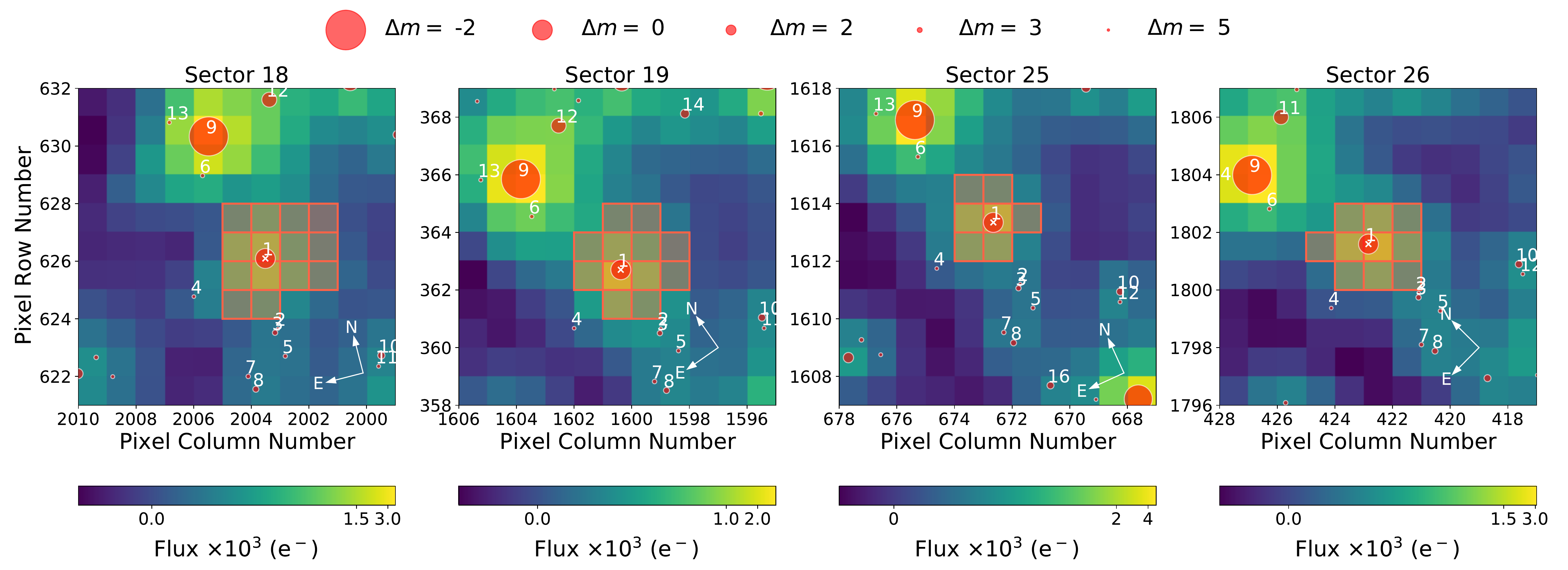}
    \caption{\tess target pixel files for all four sectors that observed the target LHS~1478. The red shaded area corresponds to the apertures used to extract the photometry. The nearest sources to the target with up to 6\,mag differences in the {\em Gaia} $G$ passband are marked with red circles. The TESS pixel scale is $\sim$ 21\,arcsec.}
    \label{fig:TFOV}
\end{figure*}

\begin{table*}
\centering
\caption{RV data and activity indicators from CARMENES VIS.}
\label{tab:carmenes_data}
\begin{tabular}{cccccc}
\hline\hline
\noalign{\smallskip}
BJD & RV & RV error & CRX & dLW & H$\alpha$ \\
-2450000 & (m\,s$^{-1}$) & (m\,s$^{-1}$) & (m\,s$^{-1}$\,Np$^{-1}$) & (m$^2$\,s$^{-2}$) & ~\\
\noalign{\smallskip}
\hline
\noalign{\smallskip}
8877.420 & -4.54 & 2.03 & 11.0 $\pm$ 17.8 & 3.3 $\pm$ 2.4 & 0.934 $\pm$ 0.003 \\
8881.348 & 1.71 & 2.40 & 2.0 $\pm$ 25.8 & -20.4 $\pm$ 4.2 & 0.943 $\pm$ 0.004 \\
8882.344 & 1.31 & 1.96 & -8.3 $\pm$ 20.0 & 3.4 $\pm$ 1.8 & 0.930 $\pm$ 0.003 \\
8883.412 & -0.62 & 1.78 & -8.0 $\pm$ 17.4 & 4.0 $\pm$ 1.9 & 0.925 $\pm$ 0.003 \\
8884.348 & -1.28 & 1.52 & 4.9 $\pm$ 14.8 & 8.2 $\pm$ 1.9 & 0.943 $\pm$ 0.003 \\
8885.346 & -0.87 & 2.33 & -3.2 $\pm$ 23.4 & 9.3 $\pm$ 2.2 & 0.968 $\pm$ 0.004 \\
8891.361 & 2.12 & 1.77 & -1.9 $\pm$ 17.9 & 4.6 $\pm$ 1.6 & 0.949 $\pm$ 0.003 \\
8893.372 & -1.07 & 1.80 & -22.7 $\pm$ 16.9 & 3.8 $\pm$ 1.9 & 0.923 $\pm$ 0.003 \\
8894.345 & -2.19 & 1.78 & -15.2 $\pm$ 14.8 & 6.0 $\pm$ 1.8 & 0.930 $\pm$ 0.003 \\
8895.364 & 2.81 & 2.54 & -50.7 $\pm$ 25.3 & -8.6 $\pm$ 2.1 & 0.925 $\pm$ 0.004 \\
8896.368 & -5.29 & 1.77 & 15.4 $\pm$ 18.2 & -4.6 $\pm$ 2.1 & 0.933 $\pm$ 0.003 \\
8897.353 & -0.43 & 2.03 & 2.0 $\pm$ 21.8 & -1.4 $\pm$ 1.9 & 0.928 $\pm$ 0.004 \\
8904.361 & -1.85 & 2.09 & 12.2 $\pm$ 18.9 & -0.7 $\pm$ 2.0 & 0.937 $\pm$ 0.004 \\
8913.365 & 0.36 & 2.24 & -26.4 $\pm$ 22.9 & 0.9 $\pm$ 3.0 & 0.919 $\pm$ 0.004 \\
8916.359 & 4.48 & 2.88 & -58.1 $\pm$ 29.3 & 4.8 $\pm$ 4.0 & 0.920 $\pm$ 0.005 \\
8918.371 & -4.55 & 2.90 & 39.2 $\pm$ 30.7 & 43.1 $\pm$ 5.2 & 0.929 $\pm$ 0.005 \\
8919.376 & -0.22 & 3.00 & 13.4 $\pm$ 31.9 & -9.4 $\pm$ 3.0 & 0.920 $\pm$ 0.005 \\
8921.328 & 1.53 & 2.07 & -0.6 $\pm$ 19.3 & -4.8 $\pm$ 2.5 & 0.916 $\pm$ 0.003 \\
8923.357 & -2.78 & 1.92 & 8.2 $\pm$ 18.2 & -2.6 $\pm$ 2.1 & 0.916 $\pm$ 0.003 \\
9035.645 & 6.60 & 2.85 & -43.2 $\pm$ 29.6 & 22.8 $\pm$ 4.5 & 0.917 $\pm$ 0.005 \\
9036.639 & -0.79 & 2.46 & -32.8 $\pm$ 25.1 & 8.7 $\pm$ 2.4 & 0.907 $\pm$ 0.003 \\
9037.633 & 2.53 & 2.19 & 12.1 $\pm$ 22.8 & 0.9 $\pm$ 1.8 & 0.920 $\pm$ 0.003 \\
9047.655 & -0.89 & 5.02 & -28.1 $\pm$ 27.8 & 3.5 $\pm$ 2.2 & 0.908 $\pm$ 0.003 \\
9049.595 & 1.42 & 2.35 & 17.0 $\pm$ 24.5 & -7.1 $\pm$ 2.3 & 0.909 $\pm$ 0.004 \\
9050.571 & -2.27 & 2.19 & -10.2 $\pm$ 19.9 & -11.9 $\pm$ 2.7 & 0.901 $\pm$ 0.003 \\
9051.580 & 3.27 & 4.09 & -38.8 $\pm$ 41.5 & -31.8 $\pm$ 5.5 & 0.922 $\pm$ 0.007 \\
9054.590 & -4.61 & 2.08 & -50.6 $\pm$ 18.2 & -0.5 $\pm$ 1.7 & 0.915 $\pm$ 0.003 \\
9055.559 & 1.72 & 1.94 & 14.1 $\pm$ 18.3 & -3.1 $\pm$ 2.3 & 0.912 $\pm$ 0.003 \\
9056.575 & 0.94 & 2.48 & 5.6 $\pm$ 23.8 & -7.8 $\pm$ 2.3 & 0.917 $\pm$ 0.004 \\
9059.567 & 6.79 & 2.78 & 27.8 $\pm$ 28.9 & -4.5 $\pm$ 2.4 & 0.932 $\pm$ 0.004 \\
9060.656 & 1.85 & 2.67 & 18.2 $\pm$ 26.4 & 1.1 $\pm$ 2.9 & 0.924 $\pm$ 0.004 \\
9061.554 & 6.87 & 2.28 & -18.4 $\pm$ 22.1 & 4.3 $\pm$ 2.6 & 0.931 $\pm$ 0.004 \\
9063.543 & 1.87 & 2.75 & 32.5 $\pm$ 29.1 & 10.0 $\pm$ 3.1 & 0.915 $\pm$ 0.005 \\
9064.535 & 3.23 & 3.40 & -18.5 $\pm$ 36.9 & 13.3 $\pm$ 3.6 & 0.920 $\pm$ 0.005 \\
9111.691 & 4.61 & 3.54 & 47.2 $\pm$ 35.5 & -4.9 $\pm$ 4.7 & 0.957 $\pm$ 0.007 \\
9112.631 & -5.92 & 2.46 & 25.0 $\pm$ 21.9 & -1.8 $\pm$ 1.8 & 0.946 $\pm$ 0.004 \\
9113.501 & 14.85 & 4.18 & 112.7 $\pm$ 42.5 & -37.7 $\pm$ 6.8 & 0.947 $\pm$ 0.007 \\
9114.403 & 3.22 & 3.41 & 67.2 $\pm$ 35.4 & -22.7 $\pm$ 5.1 & 0.946 $\pm$ 0.005 \\
9114.577 & 2.82 & 2.55 & 46.3 $\pm$ 24.7 & -2.9 $\pm$ 2.3 & 0.939 $\pm$ 0.004 \\
9115.595 & 4.73 & 3.93 & 94.9 $\pm$ 38.6 & -1.9 $\pm$ 4.4 & 0.950 $\pm$ 0.006 \\
9118.556 & 2.63 & 3.15 & 95.2 $\pm$ 27.2 & -26.8 $\pm$ 5.0 & 0.947 $\pm$ 0.005 \\
9119.480 & 5.98 & 2.49 & 10.0 $\pm$ 23.1 & -4.0 $\pm$ 3.0 & 0.934 $\pm$ 0.004 \\
9119.630 & 4.25 & 4.54 & 16.5 $\pm$ 49.0 & -5.5 $\pm$ 4.7 & 0.947 $\pm$ 0.007 \\
9120.495 & -7.31 & 1.94 & -2.1 $\pm$ 18.8 & -0.6 $\pm$ 2.0 & 0.943 $\pm$ 0.003 \\
9121.516 & 5.64 & 4.71 & 84.5 $\pm$ 49.1 & 48.3 $\pm$ 6.1 & 0.937 $\pm$ 0.008 \\
9122.504 & -5.05 & 2.59 & 42.6 $\pm$ 26.9 & 36.5 $\pm$ 4.1 & 0.935 $\pm$ 0.005 \\
9122.662 & -6.68 & 2.06 & 15.7 $\pm$ 21.2 & -3.7 $\pm$ 2.8 & 0.920 $\pm$ 0.004 \\
9126.500 & -0.99 & 2.57 & 31.7 $\pm$ 24.1 & 1.2 $\pm$ 3.2 & 0.948 $\pm$ 0.005 \\
9127.662 & -1.77 & 3.39 & 64.7 $\pm$ 32.3 & 5.6 $\pm$ 3.2 & 0.943 $\pm$ 0.005 \\
9128.527 & -1.24 & 2.44 & -32.1 $\pm$ 23.6 & 1.9 $\pm$ 1.7 & 0.941 $\pm$ 0.003 \\
9128.644 & -3.74 & 1.85 & -5.5 $\pm$ 16.6 & 3.7 $\pm$ 1.7 & 0.931 $\pm$ 0.003 \\
9132.594 & -3.04 & 2.59 & 41.0 $\pm$ 24.1 & -1.5 $\pm$ 2.4 & 0.918 $\pm$ 0.003 \\
9132.694 & 0.97 & 3.56 & 64.1 $\pm$ 35.5 & 10.3 $\pm$ 3.5 & 0.905 $\pm$ 0.006 \\
9138.394 & -1.04 & 3.77 & -69.0 $\pm$ 37.3 & -0.1 $\pm$ 3.6 & 0.943 $\pm$ 0.006 \\
9138.528 & 3.79 & 3.83 & 63.5 $\pm$ 38.4 & -1.3 $\pm$ 4.2 & 0.930 $\pm$ 0.006 \\
9138.595 & 5.45 & 3.38 & -105.6 $\pm$ 30.0 & -7.7 $\pm$ 4.0 & 0.915 $\pm$ 0.006 \\
9138.672 & 3.15 & 4.22 & -20.2 $\pm$ 44.0 & -3.6 $\pm$ 3.5 & 0.913 $\pm$ 0.007 \\
\noalign{\smallskip}
\hline
\end{tabular}
\end{table*}

\begin{table}
\centering
\caption{RV data from IRD.}
\label{tab:ird_data}
\begin{tabular}{ccc}
\hline\hline
\noalign{\smallskip}
BJD & RV & RV error \\
--2450000 & (m\,s$^{-1}$) & (m\,s$^{-1}$) \\
\noalign{\smallskip}
\hline
9120.014 & --3.25 & 3.78 \\
9120.024 & --3.05 & 5.27 \\
9123.071 & --12.62 & 3.49 \\
9123.082 & --4.52 & 3.36 \\
9152.992 & 10.43 & 5.03 \\
9156.926 & 4.11 & 3.16 \\
9157.012 & --3.44 & 3.34 \\
9188.842 & 6.78 & 6.78 \\
9188.853 & --0.39 & 5.99 \\
9189.771 & 1.52 & 3.97 \\
9189.781 & --1.77 & 4.00 \\
9189.890 & 4.69 & 3.60 \\
9189.901 & 1.49 & 3.60 \\
\noalign{\smallskip}
\hline
\end{tabular}
\end{table}

\begingroup
\renewcommand{\arraystretch}{1.5}
\begin{table}
\centering
\caption{Priors used for the \texttt{juliet} run.}
\label{tab:juliet_priors}
\begin{tabular}{lc}
\hline\hline
Parameter & Prior\tablefootmark{a}\\
\hline

$M_{\text{lc}}$\tablefootmark{b} [ppm] & $\mathcal{N} (0, 0.1)$ \\
$\sigma_{\text{lc}}$\tablefootmark{b} [ppm] & $\mathcal{LU} (0.1, 1000)$ \\
$q_{1, \text{TESS}}$ & $\mathcal{U} (0, 1)$ \\
$q_{2, \text{TESS}}$ & $\mathcal{U} (0, 1)$ \\

$q_{1, \text{LCO,RCO}}$ & $\mathcal{U} (0, 1)$ \\

$\mu_{\text{RV}}$ [m s$^{-1}$]\tablefootmark{c} & $\mathcal{N} (0, 0.2)$ \\
$\sigma_{\text{RV}}$ [m s$^{-1}$]\tablefootmark{c} & $\mathcal{LU} (10^{-4}, 15)$ \\

$P$ [d] & $\mathcal{N}(1.949, 0.01)$ \\
$T_0$ [BJD-2458000] & $\mathcal{N}(786, 0.01)$ \\
$r_1$ & $\mathcal{U}(0, 1)$ \\
$r_2$ & $\mathcal{U}(0, 1)$ \\
$a/R_{\star}$ & $\mathcal{N} (16.1,0.2)$ \\
$K$ [m s$^{-1}$] & $\mathcal{U} (0.2,10)$ \\
$e$ & Fixed(0) \\

$\sigma_{\text{GP, TESS}}$ [ppm] & $\mathcal{LU} (10^{-6}, 10^6)$ \\
$\rho_{\text{GP, TESS}}$ [d] & $\mathcal{LU} (10^{-3}, 10^3)$ \\

\hline
\end{tabular}
\tablefoot{
\tablefoottext{a}{$\mathcal{N}$: normal distribution, $\mathcal{LU}$: log-uniform distribution, $\mathcal{U}$: uniform distribution.}
\tablefoottext{b}{The same prior distributions were assumed for the photometric data, where ``lc'' stands for the light curves of TESS sectors 18, 19, 25, 26, the two nights of photometry from LCOGT, and RCO.}
\tablefoottext{c}{The same prior distributions were assumed for the RV data (CARMENES and IRD).}
}
\end{table}
\endgroup

\begingroup
\renewcommand{\arraystretch}{1.3}
\begin{table}
\caption{Posterior distributions for the instrumental parameters. The uncertainties represent the 68\,\% CI of the obtained distributions.}\label{tab:inst_parameters}
\centering
\begin{tabular}{lc}
\hline\hline
Parameter & Value\\
\hline

\noalign{\smallskip}
\multicolumn{2}{c}{\em Photometric parameters}\\
\noalign{\smallskip}
$M_{\text{S18}}$ [$10^{-6}$ ppm] & $3.97^{+57.05}_{-56.47}$ \\
$\sigma_{\text{S18}}$ [ppm] & $1.78^{+11.24}_{-1.51}$ \\
$M_{\text{S19}}$ [$10^{-6}$ ppm] & $-16.15^{+52.67}_{-50.37}$ \\
$\sigma_{\text{S19}}$ [ppm] & $1.99^{+12.69}_{-1.68}$ \\
$M_{\text{S25}}$ [$10^{-6}$ ppm] & $-75.50^{+51.51}_{-50.25}$ \\
$\sigma_{\text{S25}}$ [ppm] & $2.04^{+13.31}_{-1.75}$ \\
$M_{\text{S26}}$ [$10^{-6}$ ppm] & $-12.42^{+53.02}_{-56.05}$ \\
$\sigma_{\text{S26}}$ [ppm] & $1.91^{+11.40}_{-1.63}$ \\
$M_{\text{LCO1}}$ [$10^{-6}$ ppm] & $-115.66^{+104.37}_{-103.52}$ \\
$\sigma_{\text{LCO1}}$ [ppm] & $977.77^{+16.31}_{-30.72}$ \\
$M_{\text{LCO2}}$ [$10^{-6}$ ppm] & $-69.46^{+125.09}_{-123.14}$ \\
$\sigma_{\text{LCO2}}$ [ppm] & $856.87^{+90.00}_{-107.94}$ \\
$M_{\text{RCO}}$ [$10^{-6}$ ppm] & $10.19^{+144.64}_{-143.99}$ \\
$\sigma_{\text{RCO}}$ [ppm] & $651.31^{+250.94}_{-641.52}$ \\
$q_{1,\text{TESS}}$\tablefootmark{a} & $0.29^{+0.21}_{-0.16}$ \\
$q_{2,\text{TESS}}$\tablefootmark{a} & $0.43^{+0.34}_{-0.29}$ \\
$q_{1,\text{LCO}}$ & $0.51^{+0.26}_{-0.28}$ \\
$q_{1,\text{RCO}}$ & $0.34^{+0.31}_{-0.23}$ \\
\noalign{\smallskip}
\multicolumn{2}{c}{\em RV parameters}\\
\noalign{\smallskip}
$\mu_{\text{CARMENES}}$ [m s$^{-1}$] & $0.04^{+0.17}_{-0.16}$ \\
$\sigma_{\text{CARMENES}}$ [m s$^{-1}$] & $1.43^{+0.56}_{-0.70}$ \\
$\mu_{\text{IRD}}$ [m s$^{-1}$] & $-0.26^{+0.43}_{-0.44}$ \\
$\sigma_{\text{IRD}}$ [m s$^{-1}$] & $2.66^{+1.46}_{-1.91}$ \\

\noalign{\smallskip}
\multicolumn{2}{c}{\em GP parameters}\\
\noalign{\smallskip}
$\text{GP}_{\sigma,\text{TESS}}$ [$10^{-6}$ ppm] & $339.76^{+20.12}_{-19.07}$ \\
$\text{GP}_{\rho,\text{TESS}}$ [d] & $0.25^{+0.02}_{-0.02}$ \\
\hline
\end{tabular}
\tablefoot{
\tablefoottext{a}{Parameterization from \citet{Kipping2013}.}
}
\end{table}
\endgroup

   \begin{figure}
   \centering
   \includegraphics[width=\hsize]{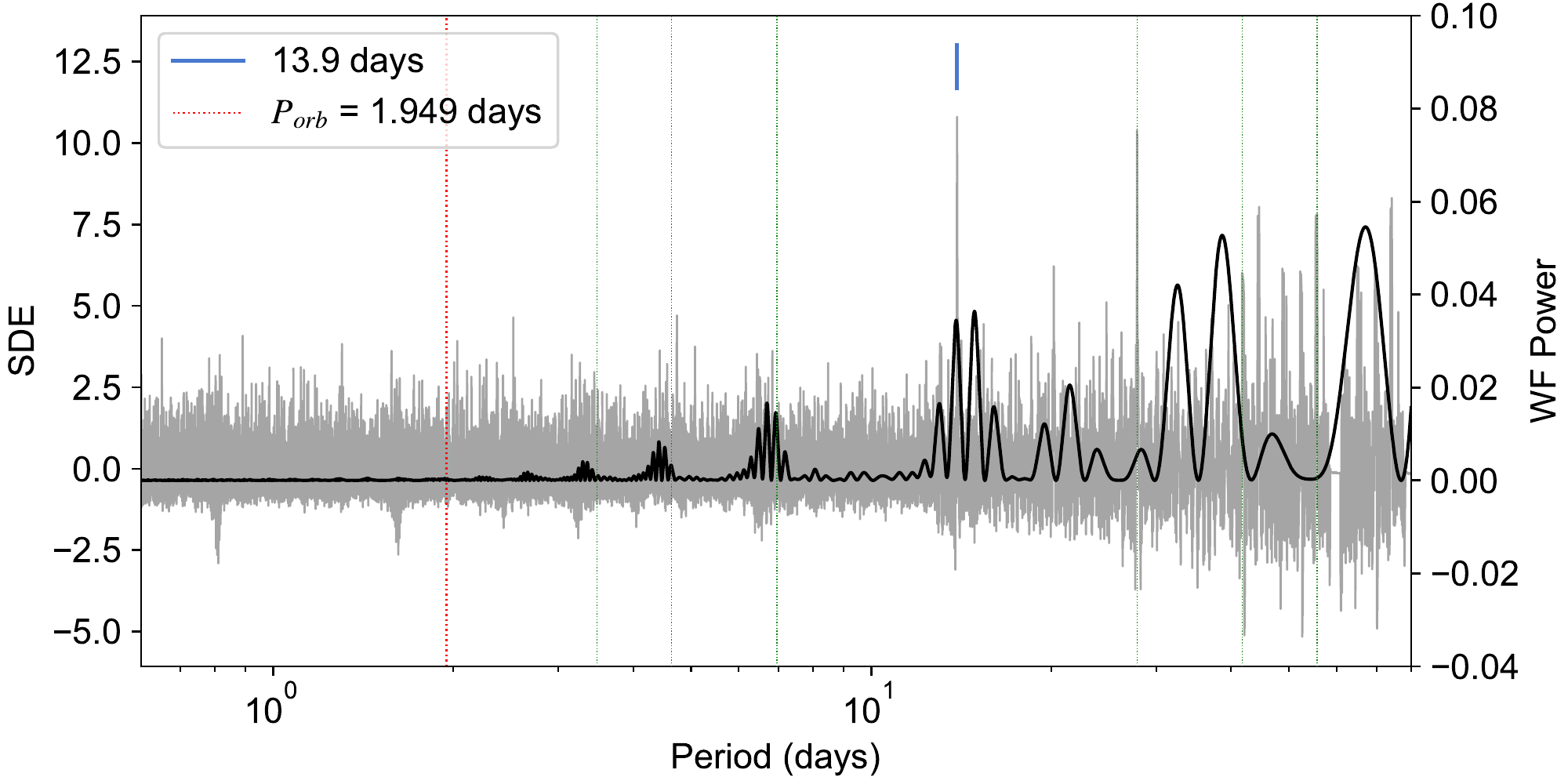}
      \caption{TLS of the \tess light curve residuals after subtracting the GP-plus-transit model. The dashed green lines are the aliases of the 13.9\,d signal. The black line represents the window function of the data.}
         \label{fig:TLS_residuals}
   \end{figure}

\end{appendix}

\end{document}